\documentclass[11pt]{article}

\usepackage[margin=1in]{geometry}
\usepackage{graphicx}            
\usepackage{amsmath,amssymb}     
\usepackage[numbers]{natbib}              
\usepackage{hyperref}            
\usepackage{authblk}             
\usepackage{booktabs}            
\usepackage{caption}             
\usepackage{xcolor}              

\makeatletter
\renewcommand\@biblabel[1]{}
\makeatother

\title{Kinematics of Single-Winged Spinning Seeds: A Study on Mahogany and Buddha Coconut Samaras }

\author[1]{Yogeshwaran G}
\author[1]{Srisha M.V. Rao}
\author[1]{Jagadeesh G}
\affil[1]{Department of Aerospace Engineering, Indian Institute of Science, Bengaluru, India}
\date{}

\begin{document}
	
	\maketitle
	
\begin{abstract}
	This study investigates the steady-state kinematics of single-winged spinning samaras and re-evaluates the simplifying assumptions commonly used in existing theoretical models. High-speed imaging was employed to quantify key parameters, including descent velocity, rotational speed, coning angle, pitching motion, and the precessional trajectory of the center of mass. The results show that all measured parameters exhibit significant temporal variation, contradicting the long-standing assumption that these quantities remain constant in steady-state flight. This variability reveals that commonly used steady-state simplifications in previous studies may overlook essential aerodynamic mechanisms governing natural samara descent. Despite this complexity, the observed sinusoidal variations in pitch, cone angle, and translational velocity, together with the nearly linear rotation rate, provide a physically grounded basis for reformulating the governing nonlinear differential equations into a simplified algebraic form. Such experimentally validated harmonic representations offer a more realistic alternative to traditional steady-state assumptions. The visualized trajectories of the center of mass, root tip, wingtip quarter-chord point, and wingtip trailing edge reveal the inherently coupled nature of samara motion and highlight the need for future experiments that capture full three-dimensional kinematics. Overall, this work advances the understanding of samara aerodynamics by emphasizing the importance of natural variability while identifying a tractable path toward more accurate modelling frameworks.
\end{abstract}
	
	\section{Introduction}
	
	\begin{figure}[!h]
		\centering
		\includegraphics[width = 0.5\textwidth]{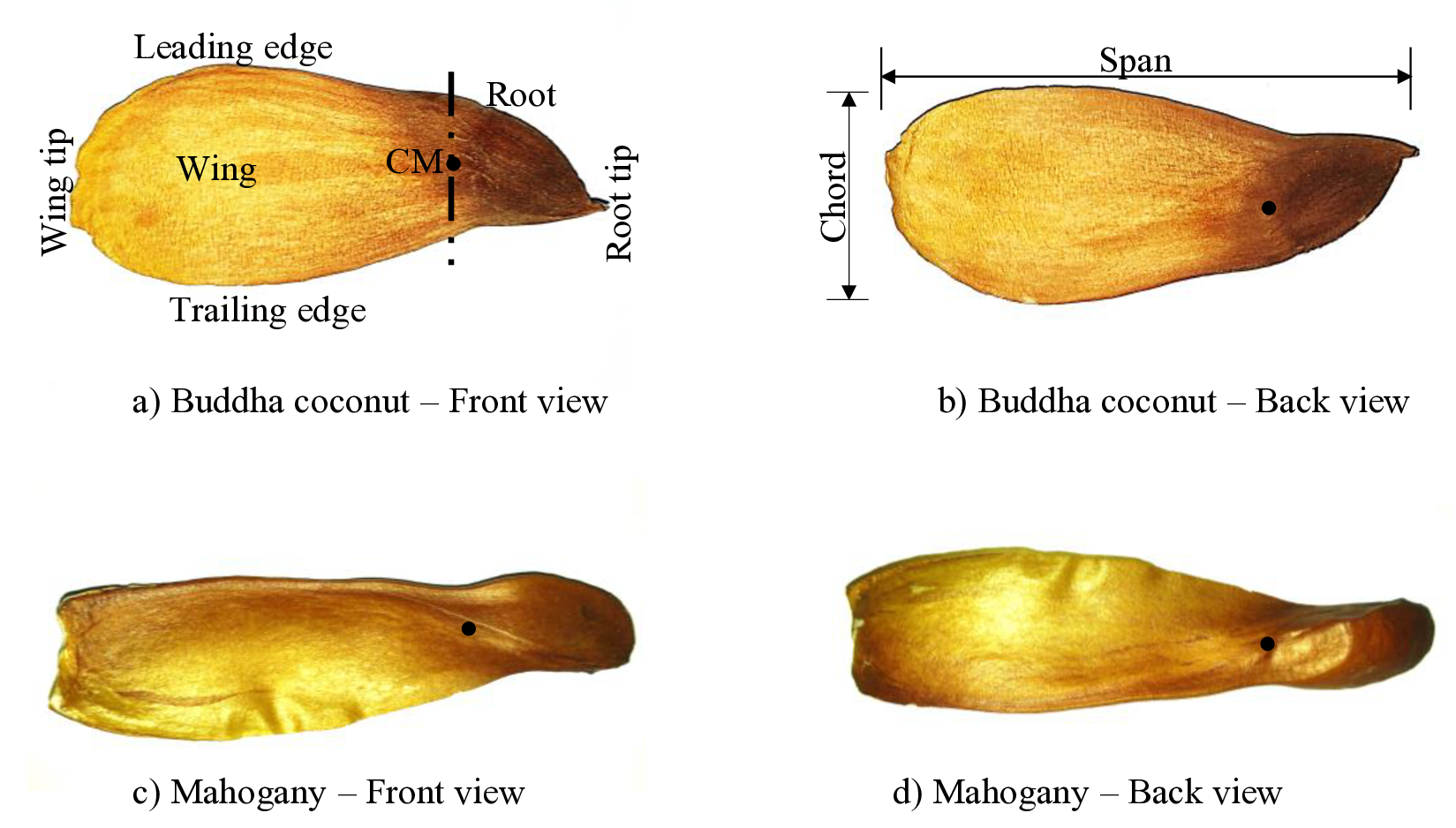}
		\caption{Front and back views of single-winged spinning samaras: (a), (b) represent the Buddha coconut samara, and (c), (d) represents mahogany samara}
		\label{samara}
	\end{figure}

	The phenomenon of objects falling through the atmosphere of the Earth has long been studied due to its complexity and broad relevance. The motion becomes particularly challenging to predict when the density of the object is close to that of air, when its mass-to-planform area ratio is very small, and when its descent is governed only by gravity and aerodynamic forces. Such trajectories, strongly influenced by geometry, mass distribution, and aerodynamic interactions, are characterized by a terminal velocity at which the gravitational and aerodynamic forces are balanced. Examples include natural systems such as feathers and artificial analogues such as a sheet of paper.

	Among natural examples of lightweight objects with small mass-to-area ratios, the seed dispersal strategy of winged seeds, or samaras, stands out as particularly remarkable \cite{NORBERG1973AutorotationFlight}. Samara utilizes its wing-like structures for aerial dispersion, taking advantage of its geometric configuration and mass distribution. This shows how nature ingeniously uses aerodynamics for efficient dispersal \cite{Green1980TheSamaras,McCutchen1977TheSamaras}. The seed is typically located in the center (in the case of multiwings) or at the corners (in the case of a single wing) of these winged structures \cite{Yogeshwaran2018OnSamaras}.  After being released from the seed pod of the parent plant, the samara undergoes a transitional phase. It then enters a steady autorotational state (also referred to as the steady state), spinning about its CM and descending gradually at a low terminal velocity. This gentle descent allows the samara to be easily caught by the oncoming wind gusts, facilitating its dispersal away from the parental plant. Fig.\ref{samara} illustrate a common instance of a single-wing spinning samara from Buddha coconut and mahogany trees.

	This natural phenomenon has inspired various bio-inspired applications in fields such as defense \cite{Crimi1988AnalysisCharacteristics}, disaster management \cite{Kellas2007TheBy}, micro air vehicles \cite{Ulrich2010PitchVehicles,Fregene2010DynamicsVehicle}, and re-entry probes \cite{Howard2009RotaryProbes}. However, fundamental questions about the interplay of various forces in relation to its kinematics remain unexplored \cite{Lentink2011Leading-EdgeOf,Schaeffer2025MapleRain}. Furthermore, even a closer examination of the kinematics of the samara reveals detailed trajectories that still require in-depth exploration \cite{Ulrich2008PlanformSamara,Lee2016NumericalSeed,Yogeshwaran2024DynamicsSamaras}. A thorough examination of the kinematics of the samara is crucial to understanding the mechanism that governs the functionality of spinning samaras.

	A central challenge in the flight of samara is to establish the relationship between geometric parameters (span, chord distribution, and mass distribution) and steady-state characteristics (descent and rotational velocities). A key step in addressing this is to model their descent using rigid-body mechanics to describe and predict their motion. This methodology provides a framework for quantifying how the geometric shape of the samara influences its steady-state kinematics. Suitable approximations of steady-state kinematics allow the governing equations of rigid-body mechanics for samaras to be greatly simplified, as noted in earlier studies.

	\begin{figure}[!h]
		\centering
		\includegraphics[width = 0.5\textwidth]{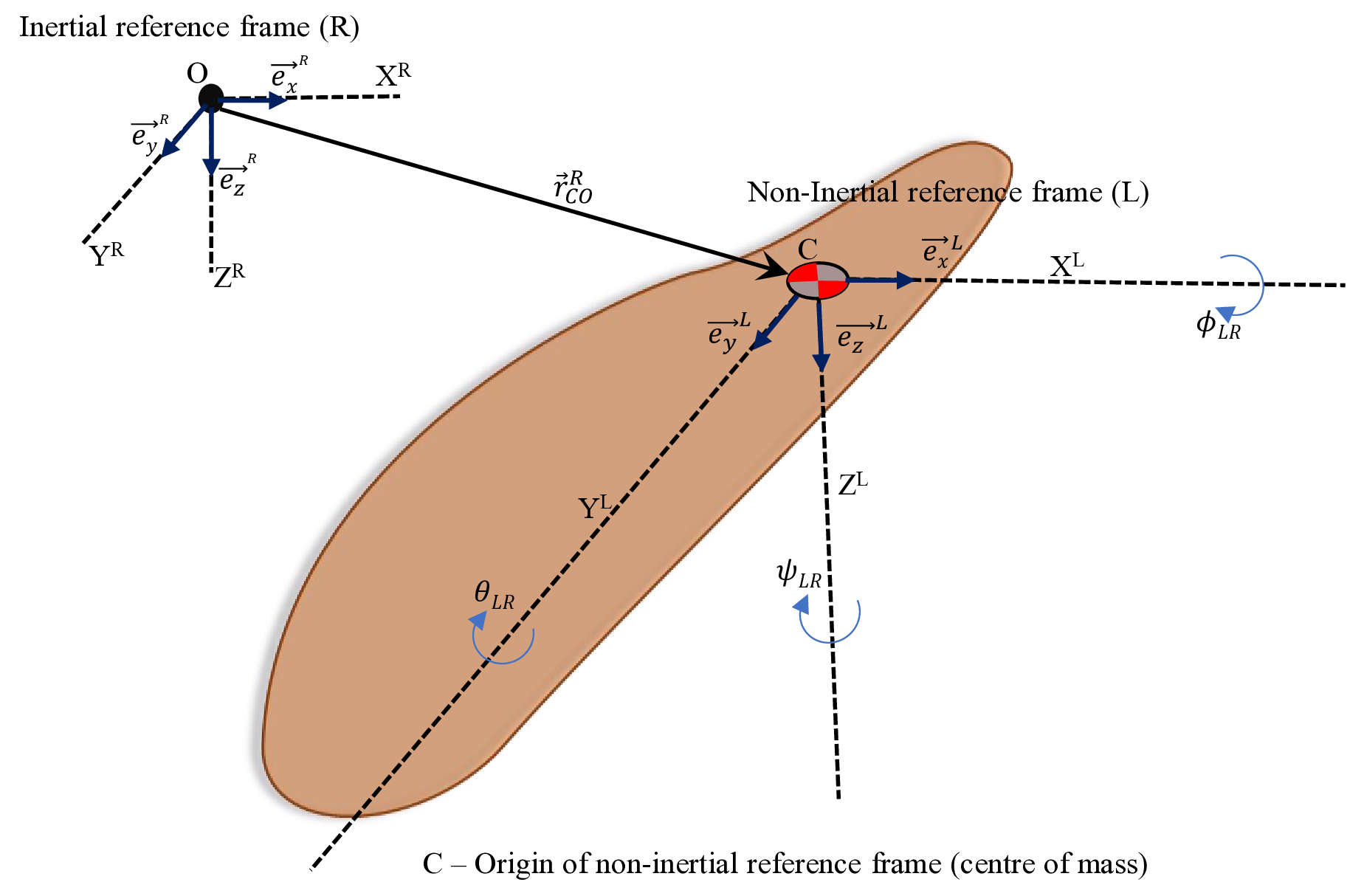}
		\caption{Analytical modelling for the single-winged spinning samara with a non-inertial reference frame attached to the CM}
		\label{analytical}
	\end{figure}

	Analytical modeling of samaras involves applying rigid body mechanics principles, specifically Newton's and Euler's laws of motion. The process is divided into two steps: deriving the equations of motion (a system of ordinary differential equations) and solving them using suitable initial conditions. To model it, two coordinate systems, the inertial and non-inertial reference frames, are defined as shown in Fig.\ref{analytical}. The inertial reference frame \(R\) is represented by the axes \(X^R, Y^R, Z^R\) and is fixed with respect to the ground. The non-inertial reference frame \(L\) is represented by the axes \(X^L, Y^L, Z^L\) and has its origin located at the CM of the samara. The non-inertial reference frame (L) is defined such that the y-axis lies along the span of the samara, the x-axis aligns with the chord, and the z-axis is perpendicular to both, positioned at the CM.
	
	Samaras are modeled as rigid bodies throughout their flight, and the equations of motion are derived using Kane’s dynamical notation \cite{Hhn2013RigidMchanics}. A superscript on the right  indicates the reference frame in which the quantity is measured  (Fig.\ref{analytical}), while a superscript on the left denotes differentiation with respect to time in the corresponding reference frame.  The orientation of the non-inertial reference frame with respect to the inertial reference frame is described by the Euler angles $(\phi_{LR}, \theta_{LR}, \psi_{LR})$, with their corresponding angular velocity components denoted by $p$, $q$, and $r$. In the steady state of the samaras, the Euler angles $\phi_{LR}$ and $\theta_{LR}$ correspond to the cone and pitch angles, respectively, while $\psi_{LR}$ represents the axis of rotation. The methodology outlined in Rigid Body Dynamics of Mechanisms by Hubert Hahn \cite{Hhn2013RigidMchanics} is used as a guiding framework to derive the governing equations for spinning samaras. The transformation matrix that connects the quantities in the inertial frame to those in the non-inertial reference frame is derived using Euler angles, as given in equation \ref{eq:ALR_def} \cite{Kellas2007TheBy}.

	
	\begin{equation}
		A^{LR} = \big[\, \mathbf{a}_1^{LR} \;\; \mathbf{a}_2^{LR} \;\; \mathbf{a}_3^{LR} \,\big]
		\label{eq:ALR_def}
	\end{equation}
	
	\begin{equation}
		\mathbf{a}_1^{LR} =
		\begin{bmatrix}
			\cos\theta_{LR}\cos\psi_{LR} \\
			\sin\phi_{LR}\sin\theta_{LR}\cos\psi_{LR}
			- \cos\phi_{LR}\sin\psi_{LR} \\
			\cos\phi_{LR}\sin\theta_{LR}\cos\psi_{LR}
			+ \sin\phi_{LR}\sin\psi_{LR}
		\end{bmatrix}
		\label{eq:a1}
	\end{equation}
	
	\begin{equation}
		\mathbf{a}_2^{LR} =
		\begin{bmatrix}
			\cos\theta_{LR}\sin\psi_{LR} \\
			\sin\phi_{LR}\sin\theta_{LR}\sin\psi_{LR}
			+ \cos\phi_{LR}\cos\psi_{LR} \\
			\cos\phi_{LR}\sin\theta_{LR}\sin\psi_{LR}
			- \sin\phi_{LR}\cos\psi_{LR}
		\end{bmatrix}
		\label{eq:a2}
	\end{equation}
	
	\begin{equation}
		\mathbf{a}_3^{LR} =
		\begin{bmatrix}
			-\sin\theta_{LR} \\
			\sin\phi_{LR}\cos\theta_{LR} \\
			\cos\phi_{LR}\cos\theta_{LR}
		\end{bmatrix}
		\label{eq:a3}
	\end{equation}

	The expressions for the time derivatives of the angular velocity components
	$(p,q,r)$ in terms of the Euler angles and their time derivatives are given by:
	\begin{align}
		p &= \dot{\phi}_{\mathrm{LR}} - \sin\theta_{\mathrm{LR}} \cdot \dot{\psi}_{\mathrm{LR}} \\[4pt]
		q &= \cos\phi_{\mathrm{LR}} \cdot \dot{\theta}_{\mathrm{LR}}
		+ \sin\phi_{\mathrm{LR}} \cdot \cos\theta_{\mathrm{LR}} \cdot \dot{\psi}_{\mathrm{LR}} \\[4pt]
		r &= -\sin\phi_{\mathrm{LR}} \cdot \dot{\theta}_{\mathrm{LR}}
		+ \cos\phi_{\mathrm{LR}} \cdot \cos\theta_{\mathrm{LR}} \cdot \dot{\psi}_{\mathrm{LR}}
	\end{align}

	Newton's second law for a rigid body of mass (m), evaluated at its CM for a spinning samara subject to the resultant external force $ \vec{F}^{\,R}$, can be expressed as:
	
	\begin{equation}
		m \, ^R\ddot{\vec{r}}^{\,R}_{\mathrm{CO}}
		\;=\;
		\vec{F}^{\,R}
		\label{eq_newtons law}
	\end{equation}

	Euler's rigid-body equation, with the inertia matrix $\widetilde{J}_c^L$ (evaluated about the CM and expressed in the non-inertial $L$ frame), is formulated using the total external moment $\vec{M}^L$.

	\begin{equation}
		\widetilde{J}_c^L \cdot \dot{\vec{\omega}}_{LR}^L
		= \vec{M}^L
		- \widetilde{\omega}_{LR}^L \cdot \widetilde{J}_c^L \cdot \vec{\omega}_{LR}^L
		\label{eq_eulers law}
	\end{equation}

	where $ \widetilde{\omega}_{LR}^L$, $\widetilde{J}_c^L$, $\vec{\omega}_{LR}^L$ and  $\dot{\vec{\omega}}_{LR}^L$ is given as
	
	\begin{equation}
		\label{omega matrix}
		\widetilde{\omega}_{LR}^L =
		\begin{bmatrix}
			0 & -r & q \\
			r & 0 & -p \\
			-q & p & 0
		\end{bmatrix}
	\end{equation}
	
	\begin{equation}
		\label{ineria matrix}
		\widetilde{J}_c^L =
		\begin{bmatrix}
			J_{cxx}^L & J_{cxy}^L & J_{cxz}^L \\
			J_{cyx}^L & J_{cyy}^L & J_{cyz}^L \\
			J_{czx}^L & J_{czy}^L & J_{czz}^L
		\end{bmatrix}
	\end{equation}

	\begin{equation}
		\vec{\omega}_{LR}^L =
		\begin{bmatrix}
			p \\
			q \\
			r
		\end{bmatrix},
		\quad
		\dot{\vec{\omega}}_{LR}^L =
		\begin{bmatrix}
			\dot{p} \\
			\dot{q} \\
			\dot{r}
		\end{bmatrix}
		\label{angualr_acc}
	\end{equation}

	The external forces and moments arise from the fluid motion around the samaras during descent, in addition to gravity, and can be modelled using the equations of fluid mechanics. The governing equations for the fluid motion around descending spinning samaras are the unsteady incompressible Navier--Stokes equation~\eqref{eq:navier_stokes}, in which $\vec{f}$ denotes the body force due to gravity, and the continuity equation~\eqref{eq:continuity}.
	
	\begin{equation}
		\label{eq:navier_stokes}
		\rho \left( \frac{\partial \vec{u}}{\partial t} + \vec{u} \cdot \nabla \vec{u} \right)
		= -\nabla p + \mu \nabla^2 \vec{u} + \vec{f}
	\end{equation}
	
	\begin{equation}
		\label{eq:continuity}
		\nabla \cdot \vec{u} = 0
	\end{equation}
	
	The governing differential equations (\ref{eq_newtons law},~\ref{eq_eulers law},~\ref{eq:navier_stokes},~\ref{eq:continuity}) that couples the geometry with steady-state kinematics can be solved using two principal approaches. The first is Direct Numerical Simulation (DNS), where rigid-body mechanics equations are integrated together with Navier-Stokes equations from the point of release until the system reaches steady state \cite{Arranz2018KinematicsSeed,Lee2016NumericalSeed}. However, this method is computationally expensive and constrained in terms of conducting parametric studies for different geometries. An alternative method is the utilization of Blade Element Momentum Theory (BEMT), but it has limitations, as it can only be applied just before autorotation starts or during its steady state \cite{Azuma1989FlightSeeds,NORBERG1973AutorotationFlight}. The second approach has been widely adopted in both fundamental research and bio-inspired applications, primarily because of its relative simplicity. In employing this method, researchers have introduced various assumptions and approximations related to aerodynamics and steady-state kinematics to solve the governing equations (Table \ref{tab:samara_analysis_summary}).
	\begin{table*}[t]
		\caption{Summary of theoretical analyses and experimental comparisons for samara studies}
		\label{tab:samara_analysis_summary}
		\scriptsize
		\centering
		
		\renewcommand{\arraystretch}{1.2}
		\setlength{\tabcolsep}{4pt}
		
		\begin{tabular}{|
				p{2.7cm}|
				p{2.0cm}|
				p{2.6cm}|
				p{3.5cm}|
				p{1.8cm}|
				p{3.0cm}|}
			\hline
			\textbf{Reference} &
			\textbf{Theory used} &
			\textbf{Geometric configuration} &
			\textbf{Key assumptions} &
			\textbf{Test domain} &
			\textbf{Accuracy comparison} \\
			\hline
			
			\cite{Crimi1988AnalysisCharacteristics}\newline
			&
			Newton--Euler,\newline
			Momentum Theory,\newline
			BET &
			Single-winged &
			Steady-state approximation with relaxed small cone and pitch angle assumptions &
			Steady state &
			10--15\% error in descent velocity and coning angle \\
			\hline
			
			\cite{Azuma1989FlightSeeds}\newline
			&
			Momentum Theory,\newline
			BET &
			Single- and multi-winged &
			Steady-state approximation &
			Steady state &
			34\% variation in descent velocity \\
			\hline
			
			\cite{Rosen1991VerticalSamara}\newline
			&
			Newton--Euler,\newline
			Momentum Theory,\newline
			BET &
			Single-winged &
			Steady-state approximation with relaxed centre-of-mass assumptions &
			Steady state &
			8.5\% descent velocity, 25\% coning angle, 17.9\% angular velocity \\
			\hline
			
			\cite{Brindejonc2007DesignSystem}\newline
			&
			Momentum Theory,\newline
			BET (RPM sweep) &
			Multi-winged &
			Steady-state approximation &
			Steady state &
			25\% error in descent velocity \\
			\hline
			
			\cite{Kellas2007TheBy}\newline
			&
			Newton--Euler,\newline
			Momentum Theory,\newline
			BET &
			Single-winged &
			Complete equation set solved without steady-state assumption &
			Pre--steady state &
			10.7\% descent velocity, 24.7\% coning angle \\
			\hline
			
			\cite{Lee2016NumericalSeed}\newline
			&
			Newton--Euler,\newline
			Navier--Stokes &
			Single-winged &
			DNS simulation &
			Transitional and \newline
			steady state &
			Terminal velocity and rotation rate matched \\
			\hline
			
			\cite{Arranz2018KinematicsSeed}\newline
			&
			Newton--Euler,\newline
			Navier--Stokes &
			Oblate spheroid nut with flat-plate wing ($Re=80$--240$)$ &
			DNS simulation &
			Transitional and \newline
			steady state &
			Agreement with prior experimental data \\
			\hline
			
			\cite{Rezgui2020ModelPerformance}\newline
			&
			Momentum Theory,\newline
			BET,\newline
			Polhamus leading-edge suction analogy &
			Single-winged &
			Steady-state approximation &
			Steady state &
			Good overall agreement \\
			\hline
			
		\end{tabular}
	\end{table*}

	\begin{figure}[!h]
		\centering
		\includegraphics[width=0.35\textwidth]{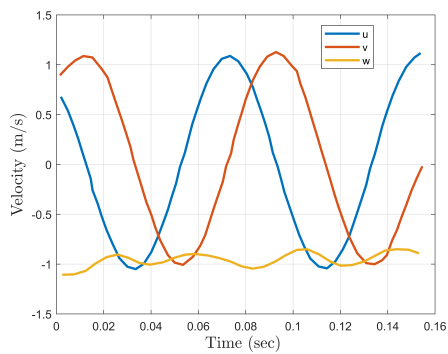}
		\caption{Velocity of the CM in the steady state with respect to the inertial reference frame \(R\), represented by u, v, and w along the \(X^R\), \(Y^R\), and \(Z^R\) directions, respectively (Data reconstructed from Varshney et al. \cite{Varshney2011TheMotion}).}
		
		\label{Varsheny}
	\end{figure}

	\begin{figure*}[!h]
		\centering
		\includegraphics[width = 0.8\textwidth]{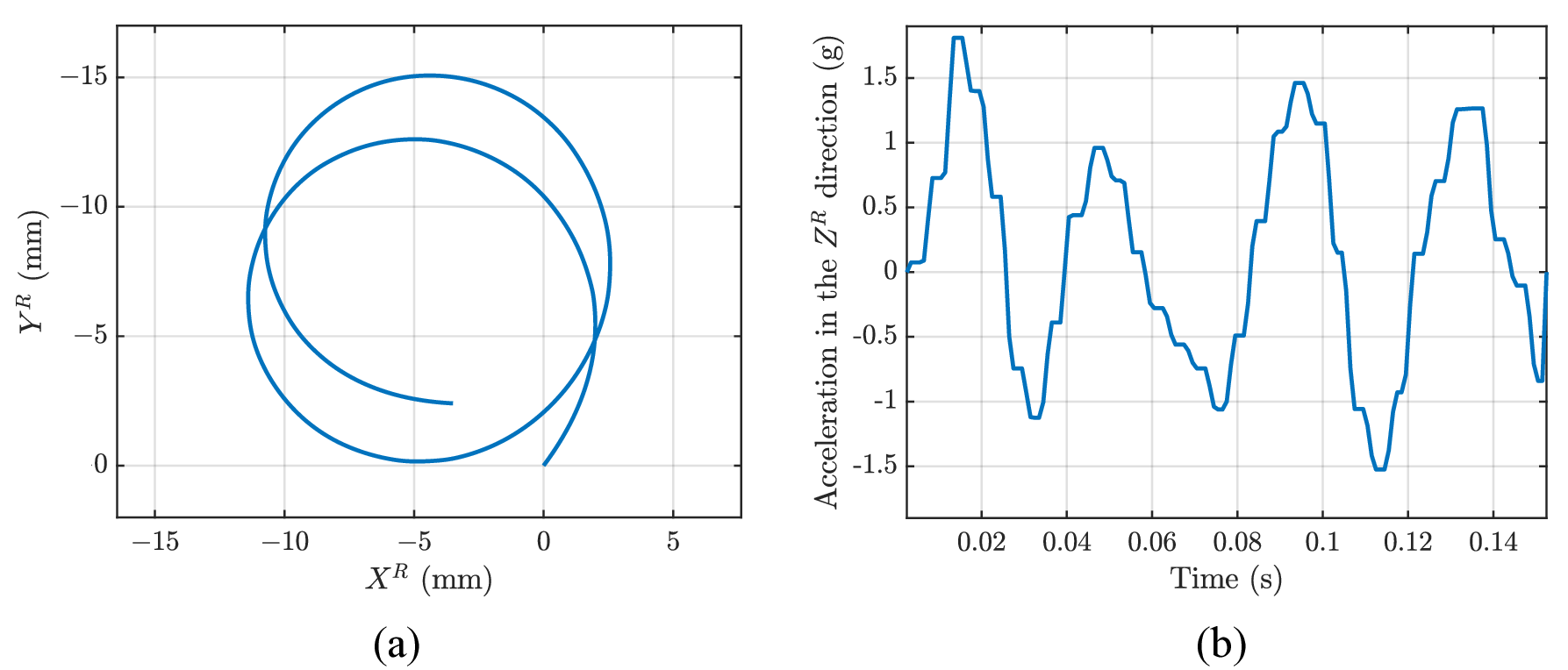}
		\caption{a) The trajectory of the CM in the X$^R$-Y$^R$ plane is obtained by integrating the u and v component velocities as depicted in Fig.\ref{Varsheny}. b) The acceleration of the CM in the downward direction is derived from the w velocity plot, as illustrated in Fig.\ref{Varsheny} from Varshney et al. \cite{Varshney2011TheMotion}.}
		\label{Trajetory_acc}
	\end{figure*}

	Pioneering experimental studies on the steady-state kinematics of natural spinning samaras were conducted by Azuma and Yasuda \cite{Azuma1989FlightSeeds}. Their results showed that the samara rotates about its CM with a constant rotational speed and a constant coning angle. In addition, they observed that the CM descends along a straight path in the direction of gravity at a constant velocity while maintaining a minimal, nearly constant pitch angle. These observations greatly simplify the governing differential equations, as the inertial terms in Newton's and Euler's laws vanish, leaving only the external forces and moments to balance. The vertical trajectory of the samara ensures that the velocity components in the $\vec{e}_x^{\,R}$ and $\vec{e}_y^{\,R}$ directions are zero, while the constant descent velocity eliminates the $\vec{e}_z^{\,R}$ component of acceleration (Fig.\ref{analytical}). Thus, in equation (\ref{eq_newtons law}) the inertial terms vanish, resulting in a balance of external forces between aerodynamics and gravity. Similarly, the constant cone angle ($\phi_{LR}$) and the pitch angle ($\theta_{LR}$) imply that $\dot{\phi}_{LR}$ and $\dot{\theta}_{LR}$ vanish, leaving only the term $\dot{\psi}_{LR}$. Furthermore, under the small-angle approximation for cone and pitch angles, the angular rates $p$ and $q$ reduce to zero. In addition, the constancy of the remaining angular rate $r$ eliminates the angular acceleration term in equation~\ref{eq_eulers law}. This assumption, which greatly simplifies the governing differential equations, is here referred to as the steady-state approximation in the motion of samaras.

	Following earlier works and reinforced by these observations, subsequent studies have widely employed the steady-state approximation, often with minor modifications. This framework has been used to solve the governing equations using BEMT, as summarized in Table~\ref{tab:samara_analysis_summary}. Crimi~\cite{Crimi1988AnalysisCharacteristics} relaxed the assumption of small cone and pitch angles for the design of a samara-inspired delivery system, while Azuma et al. \cite{Azuma1989FlightSeeds}, Brindejonc et al. \cite{Brindejonc2007DesignSystem} and Rezgui et al. \cite{Rezgui2020ModelPerformance} applied the steady-state approximation to analyze natural samaras. Rosen and Seter \cite{Rosen1991VerticalSamara} extended the framework by assuming constant center-of-mass velocity in all directions. A notable contribution by Kellas~\cite{Kellas2007TheBy} simulated the motion from the pre-steady state, where autorotation begins without relying on steady-state approximations. Furthermore, the study reported non-zero $\vec{e}_x^{,R}$ and $\vec{e}_y^{,R}$ components of the CM that varied with time, whereas the other parameters were consistent with steady-state approximations. Despite these efforts (Table~\ref{tab:samara_analysis_summary}), the accuracy of predictions compared to experimental measurements varies considerably, primarily due to the assumptions invoked in steady-state modeling. This discrepancy raises important questions regarding the true trajectory of samaras, even though the simplified approximation greatly reduces the complexity of the governing equations.

	On the other hand, more recent experimental studies on the kinematics of natural samaras and their models, conducted by Varshney et al. ~\cite{Varshney2011TheMotion} and Ulrich et al.~\cite{Ulrich2008PlanformSamara}, reported results that contradict earlier findings.  Varshney et al. \cite{Varshney2011TheMotion} investigated the three-dimensional trajectory of the CM of maple samaras along with their Euler angles. The experimental results revealed that the velocity of the CM in the horizontal plane (u and v) follows a periodic profile with a magnitude close to the descent velocity as shown in Fig.\ref{Varsheny}. The cone angle displayed periodic variation, and the pitch angle appeared negligible, consistent with Azuma et al. \cite{Azuma1989FlightSeeds}. The trajectory of the CM of samara can be derived from their corresponding velocity plots by integrating the velocity plots and considering the starting point as the origin. The trajectory in the x-y plane is depicted in the Fig.\ref{Trajetory_acc}(a). As expected from the periodic plot, the CM follows a circular path with diameter of 14 mm (approx.) as it descends. This motion is termed precession, where the samara spins about its CM, and the CM spins about the vertical axis, confirming a non-vertical trajectory for the CM. 
	
	Upon closely examining, the descent velocity of the samara as a function of time, as reported by Varshney et al. \cite{Varshney2011TheMotion} and Lee et al. \cite{Lee2016EffectPalmatum} in the steady state, a unique feature related to the concept of terminal velocity is observed. It is evident that the descent velocity, as a function of time in these two studies, is not constant and varies rapidly. The acceleration plot, calculated as a function of time for Varshney et al. \cite{Varshney2011TheMotion}, is shown in the Fig.\ref{Trajetory_acc}(b). The variation in acceleration of the CM ranges from -2g to 2g for Varshney et al. \cite{Varshney2011TheMotion} (as in Fig.\ref{Trajetory_acc}(b)) and -0.2g to 0.2g for the case of Lee et al. \cite{Lee2016EffectPalmatum}. This observation appears to contradict the concept of terminal velocity for these spinning samaras, or the variation in acceleration may be considered minimal and negligible for the governing differential equation, which warrants further investigation. Complementing these findings, Ulrich and Pines ~\cite{Ulrich2008PlanformSamara} examined 3D-printed samara models and observed a more pronounced precession, resulting in a clear helical trajectory. In this case, both pitch and cone angles exhibited periodic variation, and the pitch and roll rates were found to be neither small nor constant-factors that must be carefully considered when applying simplified governing equations.
	\begin{table*}[t]
		\centering
		\caption{Geometric and steady-state parameters of different single-wing samara models and prototypes reported in the literature}
		\label{jlab1}
		\footnotesize
		
		\setlength{\tabcolsep}{8pt}      
		\renewcommand{\arraystretch}{1.25}
		
		\begin{tabular}{|
				p{3.2cm}|
				p{3.0cm}|
				p{3.2cm}|
				p{3.2cm}|
				p{2.5cm}|}
			\hline
			\textbf{References} &
			\cite{Azuma1989FlightSeeds} &
			\cite{Ulrich2008PlanformSamara} &
			\cite{Varshney2011TheMotion} &
			\cite{Lee2016NumericalSeed} \\
			\hline
			
			\textbf{Model / Prototype} &
			Prototypes &
			Models &
			Prototype &
			Prototypes \\
			\hline
			
			Span (mm) &
			24.52 &
			162 &
			-- &
			14 \\
			\hline
			
			Mass / Area (g\,/mm$^{2}$) &
			2.11 &
			4.16 &
			2.72 &
			2.35 \\
			\hline
			
			Radius of precession (mm) &
			-- &
			166.66 &
			$\approx 14$ &
			-- \\
			\hline
			
			Conning angle ($^\circ$) &
			21.77 &
			$-10$ to $10$ &
			$-20$ to $-30$ &
			9 to 13 \\
			\hline
			
			Pitch angle ($^\circ$) &
			$\approx 1$--2 &
			$-10$ to $10$ &
			$\approx 0$ &
			$-4$ to 1 \\
			\hline
			
			Descent velocity (mm/s) &
			980 &
			1600 &
			940 &
			970 \\
			\hline
		\end{tabular}
		\label{table1}
		
	\end{table*}

	DNS simulations conducted by Arraz et al. \cite{Arranz2018KinematicsSeed} and Lee and Choi \cite{Lee2016NumericalSeed,Lee2018ScalingVelocity} partly confirmed similar behavior, particularly in the variation of pitch and cone angles, with the CM undergoing precession. Lee \cite{Lee2016NumericalSeed} observed periodic variation in cone angle from 9$^\circ$ to 14$^\circ$ and in pitch angle in the range of -4$^\circ$ to 3$^\circ$. Lee \cite{Lee2016NumericalSeed} also demonstrated that each samara undergoes rotation in both directions and showed that samaras with a clockwise direction have a lower descent velocity than anticlockwise ones for different orientations. In conclusion, the motion of samaras in the steady state is not straightforward and entails complex trajectories, as evident from the  results of the literature. The exact motion of the samara has yet to be explored, and further study is needed to develop approximations that can help simplify the governing differential equations.

	One potential reason for the discrepancy between the initial and more recent results can be attributed to the span of the samaras chosen for the studies. Azuma and Yasuda ~\cite{Azuma1989FlightSeeds} used samaras with a span of 20--30 mm, where variations in kinematic parameters were likely negligible, whereas Ulrich and Pines~\cite{Ulrich2008PlanformSamara}, and Kellas~\cite{Kellas2007TheBy} employed samara models with spans exceeding 160 mm, which resulted in substantial variations in all steady-state kinematic parameters. Table~\ref{table1} illustrates how these differences in scale contributed to the initial misconceptions in determining the trajectory of the samara.
	
	Building on the insights from previous studies, the present work investigates the kinematics of natural samaras in the steady state for spans of 60-100 mm, a scale range not addressed in existing literature. The primary objective is to examine whether the steady-state approximation is truly valid at this intermediate scale and to assess how the associated kinematic parameters vary. To this end, controlled drop experiments were conducted in the laboratory using high-speed camera measurements, and steady-state parameters were extracted using suitable image processing techniques. The study focused on five key parameters: descent velocity, coning angle, rotational velocity, wingtip pitch, and precession, which were clearly defined and systematically measured across multiple cases. Based on the results of the present study, a more realistic and aerodynamically consistent steady-state formulation is proposed. Finally, the study concludes by presenting the theoretical trajectories of samara motion by tracking the CM, root tip, wingtip quarter-chord point, and wingtip trailing edge, as derived from the present findings.

	\section{Materials and methods}
	
	\subsection{Measurement of geometric properties of the samara}
	\label{startsample}
	The mahogany (Swietenia mahagoni) and Buddha coconut (Pterygota alata) samaras were collected from the Indian Institute of Science, Bangalore, India, and used for experimental investigations. Three samaras of each species were selected, labeled MD1-MD3 for mahogany and BD1-BD3 for Buddha coconut. The platform area and span of the chosen samaras are listed in Table~\ref{geometric_kineamtics}. The mass of each samara was measured using Sartorius weighing equipment with a precision of 0.001 mg, as also reported in the same table.

	\begin{table}[!htbp]
		\centering
		\caption{Geometric properties of the chosen samaras}
		\label{geometric_kineamtics}
		
		\setlength{\tabcolsep}{6pt}
		\renewcommand{\arraystretch}{1.25}
		\footnotesize
		
		\begin{tabular}{|
				p{0.6cm}|
				p{0.6cm}|
				p{0.6cm}|
				p{2cm}|
				p{2.3cm}|}
			\hline
			\textbf{Ref. No.} &
			\textbf{Mass}\newline (g) &
			\textbf{Span}\newline (mm) &
			\textbf{Planform area}\newline (mm$^{2}$) &
			\textbf{Mass / Area}\newline ($\mathrm{g/mm^{2} \times 10^{-2}}$) \\
			
			\hline
			
			\multicolumn{5}{|c|}{\textbf{Mahogany samples}} \\
			\hline
			MD1 & 0.537 & 90.7 & 1761 & 0.0304 \\
			\hline
			MD2 & 0.481 & 91.5 & 1821 & 0.0264 \\
			\hline
			MD3 & 0.390 & 76.9 & 1353 & 0.0288 \\
			\hline
			
			\multicolumn{5}{|c|}{\textbf{Buddha coconut samples}} \\
			\hline
			BD1 & 0.825 & 80.1 & 1911 & 0.0431 \\
			\hline
			BD2 & 0.871 & 72.5 & 1668 & 0.0522 \\
			\hline
			BD3 & 0.763 & 77.7 & 1820 & 0.0419 \\
			\hline
		\end{tabular}
		
	\end{table}

	\subsection{Hold and release mechanism}
	
	\begin{figure}[!h]
		\centering
		\includegraphics[width = 0.5\textwidth]{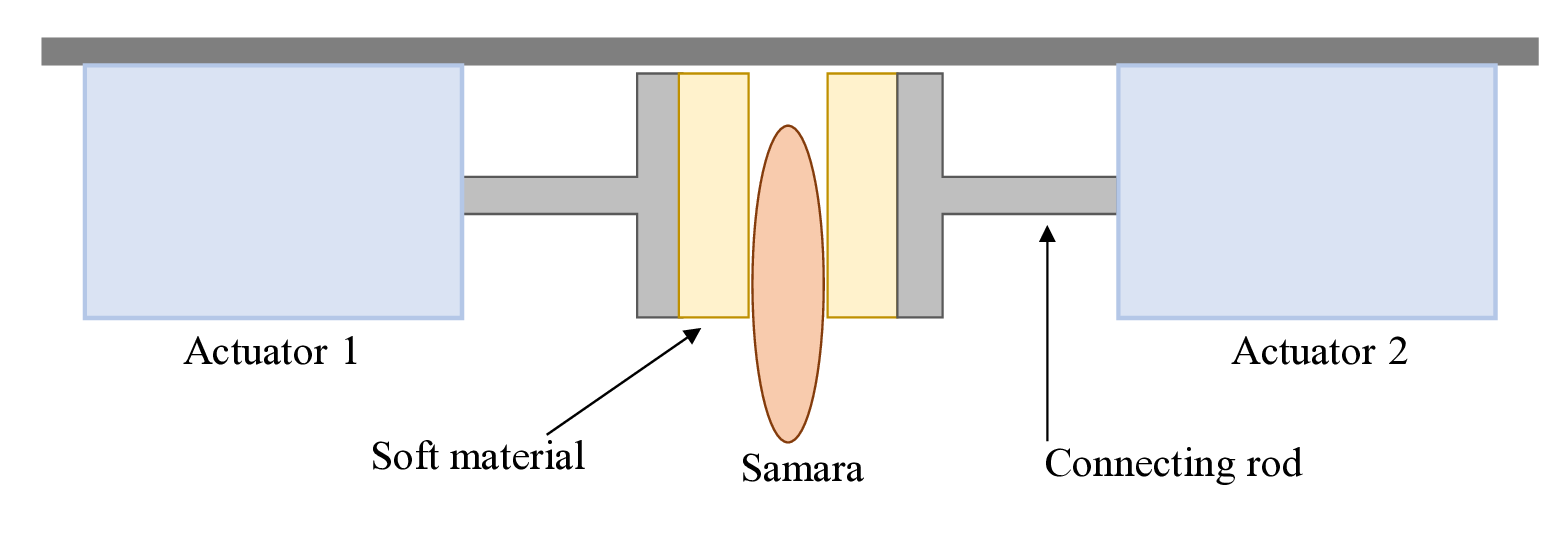}
		\caption{The schematic of the hold and release mechanism}
		\label{release mechanism}
	\end{figure}

	A hold-and-release mechanism was designed to consistently release the samara in a chosen orientation with high repeatability. The setup consists of two linear actuators that move symmetrically in opposite directions from a central point (Fig.\ref{release mechanism}), each fitted with a sponge to grip the samara without damage. Upon activation, the actuators retract simultaneously, releasing the samara from rest. Unlike the clap-and-release systems reported in earlier studies ~\cite{Ulrich2008PlanformSamara,Lee2016EffectPalmatum}, this mechanism ensures controlled and repeatable experiments through uniform actuator motion.

	\subsection{Experimental setup for drop studies}
	
	\begin{figure*}[!t]
		\centering
		\includegraphics[width=0.95\textwidth]{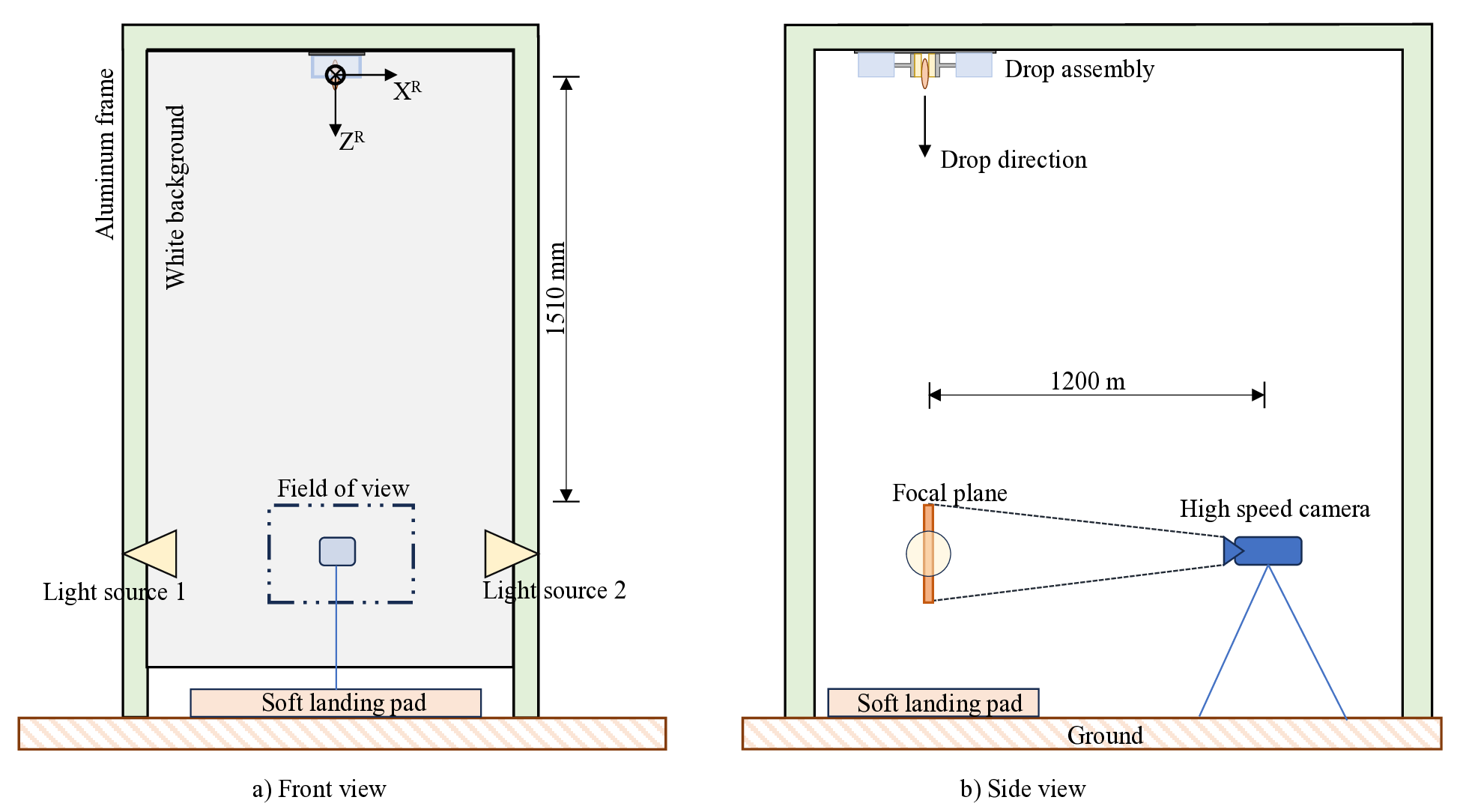}
		\caption{Schematic of the experimental setup for drop studies.}
		\label{drop assembly}
	\end{figure*}

	The experimental setup for capturing the steady-state behaviour of spinning samaras consisted of a high-speed camera, light sources, a white background, and the hold-and-release mechanism. As shown in Fig.\ref{drop assembly}, the white background was mounted on an aluminium frame with light sources positioned on either side, while the hold-and-release mechanism was placed at the top. A Phantom v310 high-speed camera operating at 1000 fps and equipped with a Nikon 40 mm lens was employed at a resolution of 1280$\times$800 pixels, with the camera and release system triggered simultaneously to ensure consistent timing across all runs. The origin of the experimental setup was defined at the midpoint of the bottom edge of the hold-and-release assembly, with the $Z^R$-axis pointing downward, corresponding to the inertial reference frame(R) shown in Fig.\ref{drop assembly} (a). The field of view was set to 514.50 mm $\times$ 321.45 mm, with its top edge located at $Y_R = 1510$ mm from the origin of the coordinate system.  A planar checkerboard pattern was used to examine the presence of any geometric distortion in the captured images caused by the camera lens. The measured square sizes varied by less than one pixel across the field of view, indicating negligible lens distortion (less than 0.1\% relative error). Therefore, no explicit distortion correction was applied to the captured images. As the samara enters the field of view from the release point, it is not always possible for it to pass precisely through the mid-plane of the focal field. Instead, it typically falls either ahead of or behind the focal plane within the field of view. This results in a magnification of the apparent size of the samara when it moves ahead of the focal plane and a demagnification when it falls behind it. Such motion primarily occurs during the transition phase, which is random in nature, and is influenced by the precession of the samara that causes out-of-plane movement. Although the focal plane and release position remain fixed, this type of error is unavoidable. The magnification or demagnification values were estimated from the projected span of the samara within the field of view, and the observed mean variation in span was found to be less than 4\% across all the runs considered.
	
	\subsection{Drop orientations}

	\begin{figure}[!h]
		\centering
		\includegraphics[width = 0.5\textwidth]{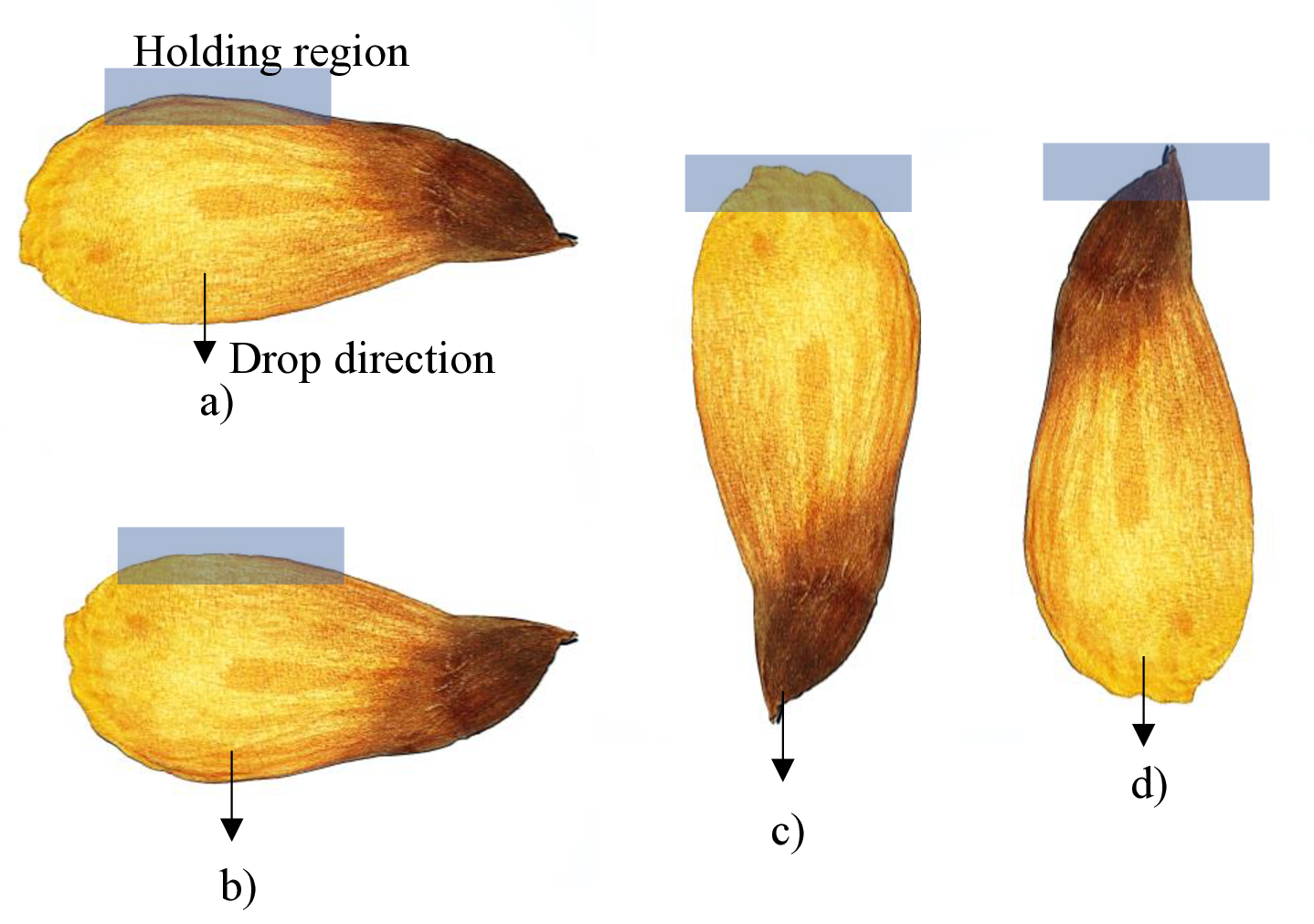}
		\caption{Initial drop orientations considered for the analysis}
		\label{drop_orinetaion}
	\end{figure}
	
	In the steady state, samaras can rotate in either direction (clockwise or anticlockwise), leading to two possible steady-state solutions. The parameters associated with these steady states vary significantly depending on the direction of rotation. The direction of rotation of the samara is determined by its initial orientation during release \cite{Lee2016NumericalSeed,Lee2016EffectPalmatum}. Furthermore, the transition length of the samara also depends on the initial orientation set in the drop assembly. To examine whether each samara exhibits two steady-state solutions and to capture its kinematics in both rotational directions, four different release orientations were considered in the present study. These orientations correspond to holding the samara at the wingtip, root tip, leading edge, and trailing edge, as shown in the Fig.\ref{drop_orinetaion}. For each samara and each release orientation, three independent runs were conducted to characterize the kinematics comprehensively.

	\subsection{Post-processing the obtained images}
	
	\begin{figure}[!h]
		\centering
		\includegraphics[width = 0.5\textwidth]{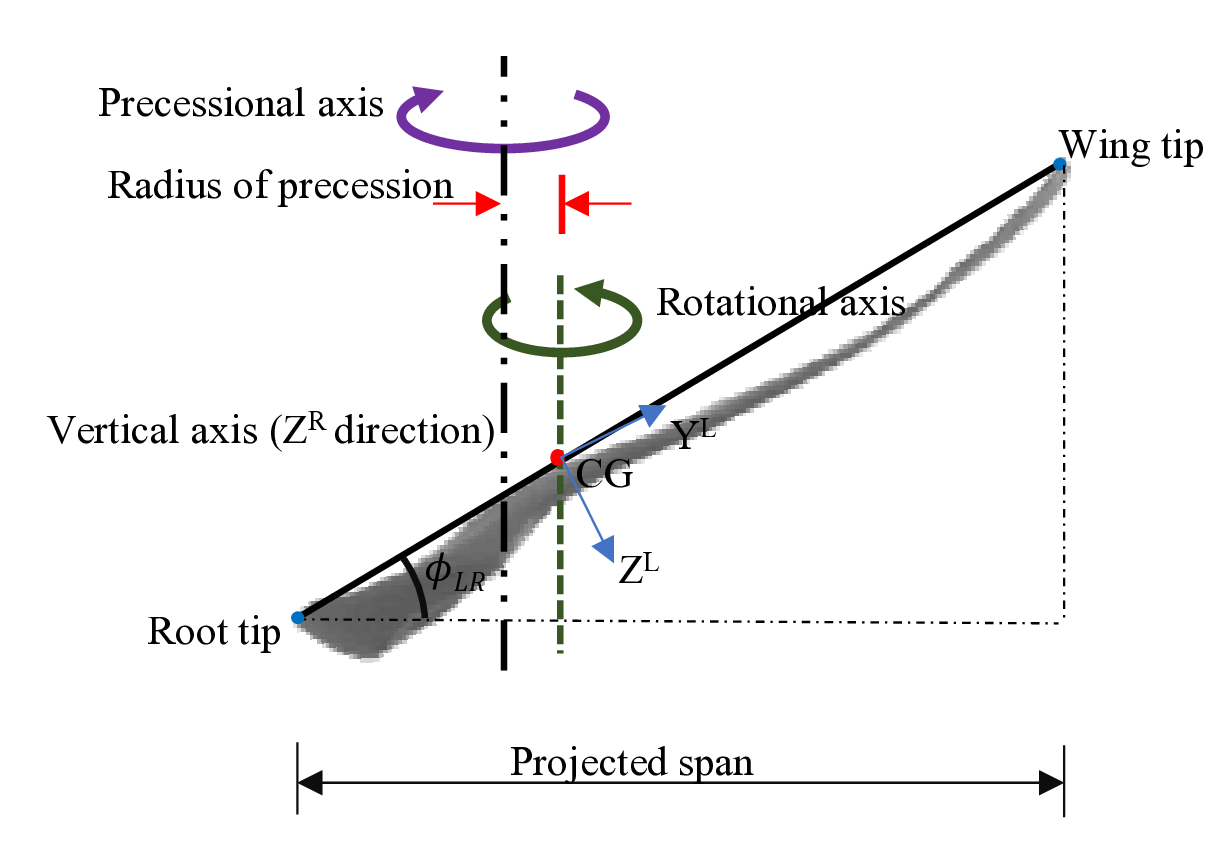}
		\caption{Schematic illustrating the calculation of steady-state parameters for the right-extreme orientation}
		\label{steady_state_parameters}
	\end{figure}

	The post-processing of the captured images was carried out using custom-developed MATLAB scripts. Initially, background subtraction was performed on all frames to isolate the samara from the background. To visualize the motion, image overlays were generated by superimposing subsequent frames onto the first instance of the samara. Steady-state kinematic parameters, including coning angle, rotational rate, precession, wingtip pitch, disk diameter, and descent velocity, were extracted from the captured images. Firstly, the projected area, defined as the number of pixels occupied by the samara in each frame, and the projected span, defined as the distance between the wingtip and root tip along the X$^{R}$ direction, were quantified and plotted as functions of time. The local minimum in the projected area-time plot and the local maximum in the projected span-time plot correspond to the leftmost and rightmost extreme orientations of the samara, respectively. These orientations were then identified on a case-by-case basis, depending on the most distinct periodic trend observed in either the projected area-time or projected span-time plots, and were subsequently used to define other kinematic parameters. The coning angle ($\phi_{LR}$) was defined as the angle between the wingtip and the root tip, measured at the identified extreme orientations (Appendix A).  The angular velocity was calculated based on the time taken for the samara to travel between these extreme orientations. The pitch of the wingtip (per half roataion) was estimated by tracking the displacement of the wingtip over half rotation. Accurately tracking the CM from the projected images proved challenging. However, based on prior observations, it was determined that the CM lies approximately 30\% of the distance from the root tip along the line connecting the wingtip and root tip. Thus, the CM position for left or right extreme orientation was estimated along this line. The vertical axis in the $Z^{R}$ direction, which serves as the precession axis of the samara, was determined as the midpoint between the leftmost and rightmost extremities of the samara corresponding to its left and right orientation images. The radius of precession was calculated as the horizontal distance between the CM and the vertical axis in the $Z^{R}$ direction. The trajectory of the bottommost point of the samara was extracted across all frames. The descent velocity was then obtained by differentiating this trajectory with respect to time, followed by applying a moving average filter to minimize noise in the velocity measurements. The orientation of the samara at its leftmost positions is shown in Fig.\ref{steady_state_parameters} along with other kinematic parameters used for measurement.

	\section{Results and Discussions}
	
	The results derived from the drop experiments are presented and discussed in two parts. First, the variation of the kinematic parameters for four selected cases is examined, followed by a detailed analysis of the overall variation of these parameters across all samples. For the four representative cases, the MD2 and BD1 samaras were chosen, as both exhibited clockwise and anti-clockwise rotations under different release orientations. This is followed by a statistical analysis of each steady-state parameter to assess the validity and limits of the steady-state approximation for the spinning samaras.

	\subsection{Case-wise analysis of kinematic variations}

	\begin{figure*}[!h]
		\centering
		\includegraphics[width = 0.7\textwidth]{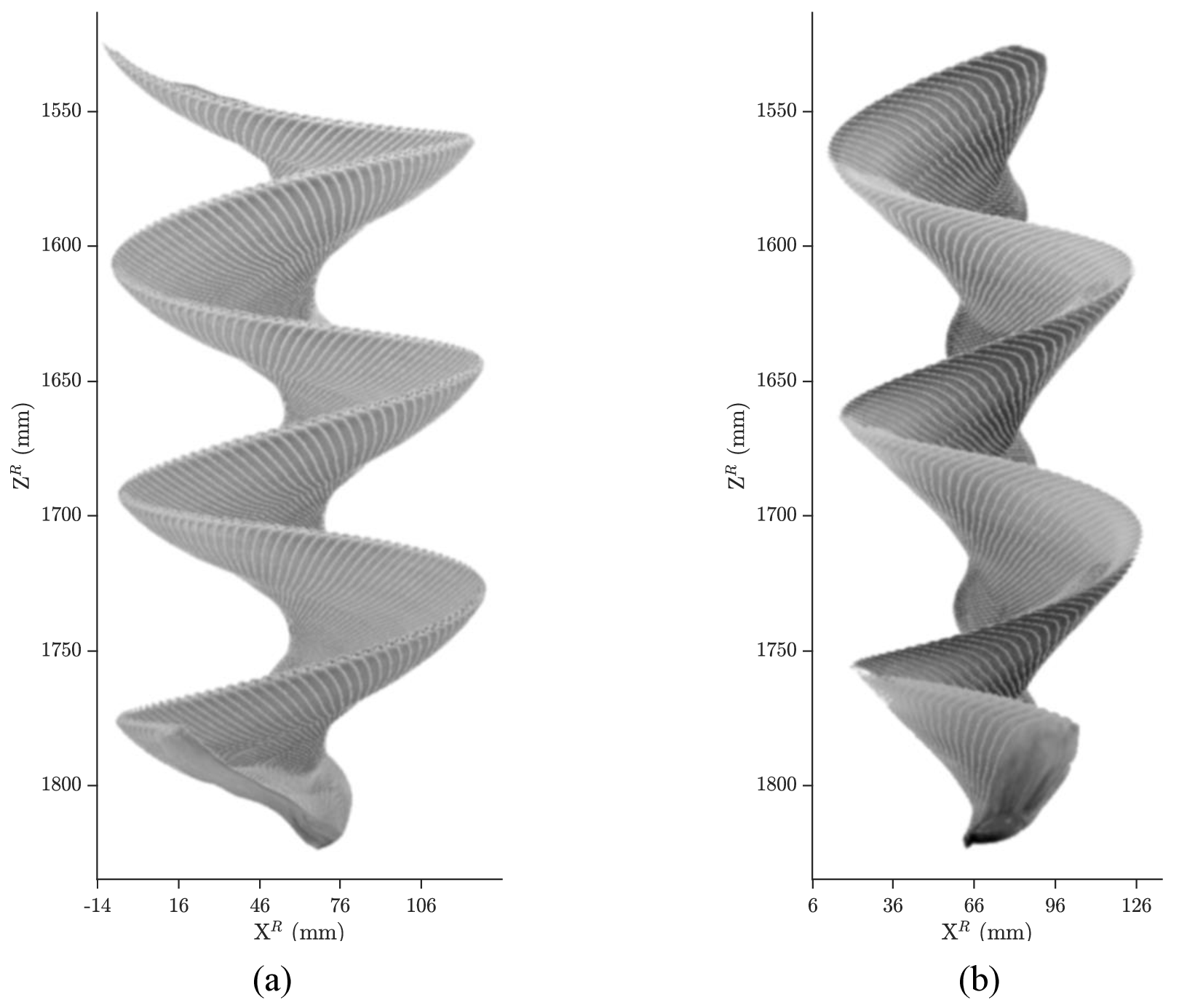}
		\caption{Image overlays at steady state for (a) Mahogany samara MD2 - anticlockwise rotation (run 19), and (b) Buddha coconut samara BD1 - clockwise rotation (run 11).}
		\label{Mahogany_Buddha_overlays}
	\end{figure*}

	\begin{figure*}[!h]
		\centering
		\includegraphics[width = 0.8\textwidth]{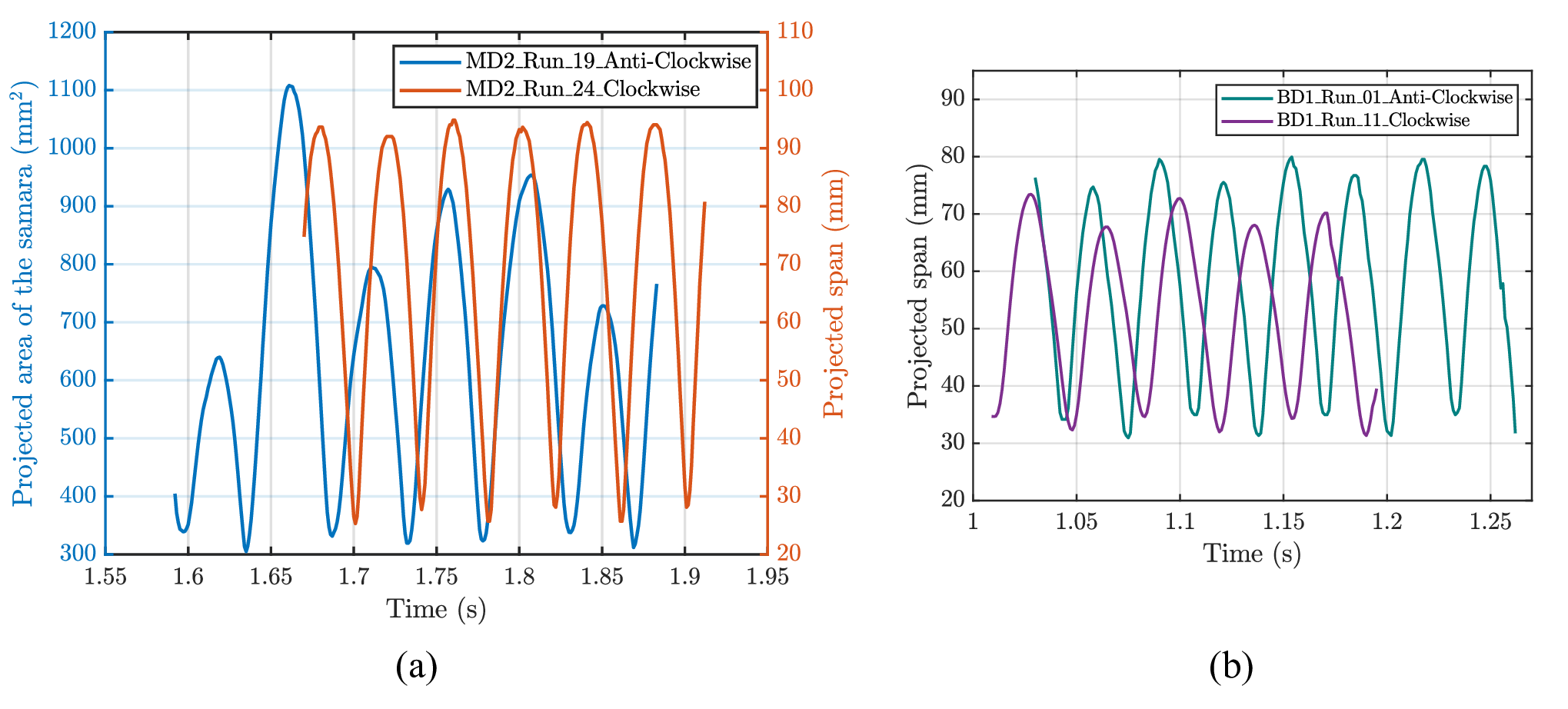}
		\caption{Projected area vs.\ time and projected span vs.\ time for (a) Mahogany samara MD2 - anticlockwise rotation (run 19) and clockwise rotation (run 24), and (b) Buddha coconut samara BD1 - clockwise rotation (run 11) and anticlockwise rotation (run 01).}
		\label{projected_lengths_width}
	\end{figure*}

	\begin{figure*}[!h]
		\centering
		\includegraphics[width = 0.8 \textwidth]{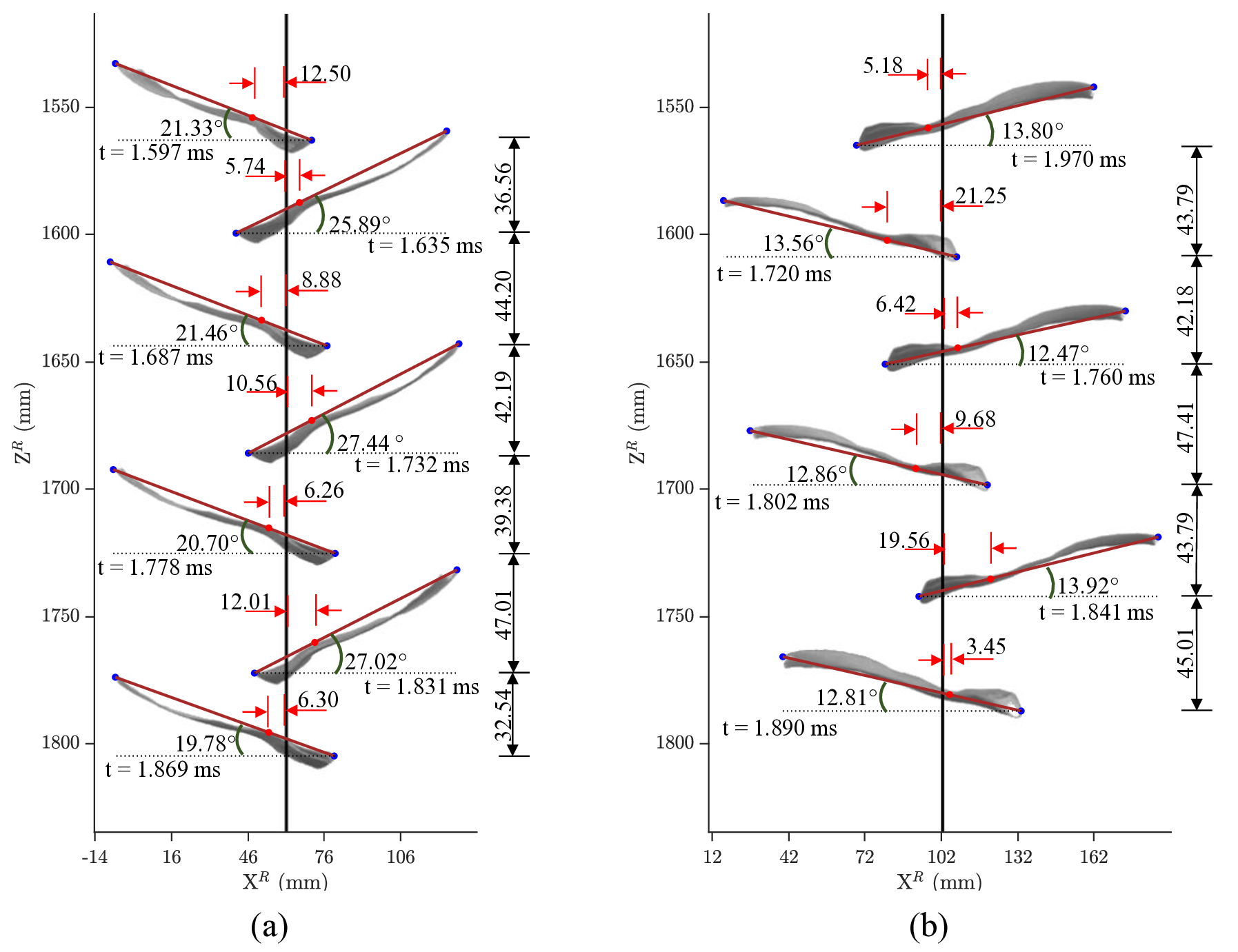}
		\caption{Measured steady-state parameters for the mahogany samara: (a) MD2 -anticlockwise rotation (run 19) and (b) MD2 - clockwise rotation (run 24). The wingtip pitch (per half rotation) and the radius of precession are reported in millimetres (mm).}
		\label{Cone_mahogany_overlays}
	\end{figure*}

	\begin{figure*}[!h]
		\centering
		\includegraphics[width = 0.7\textwidth]{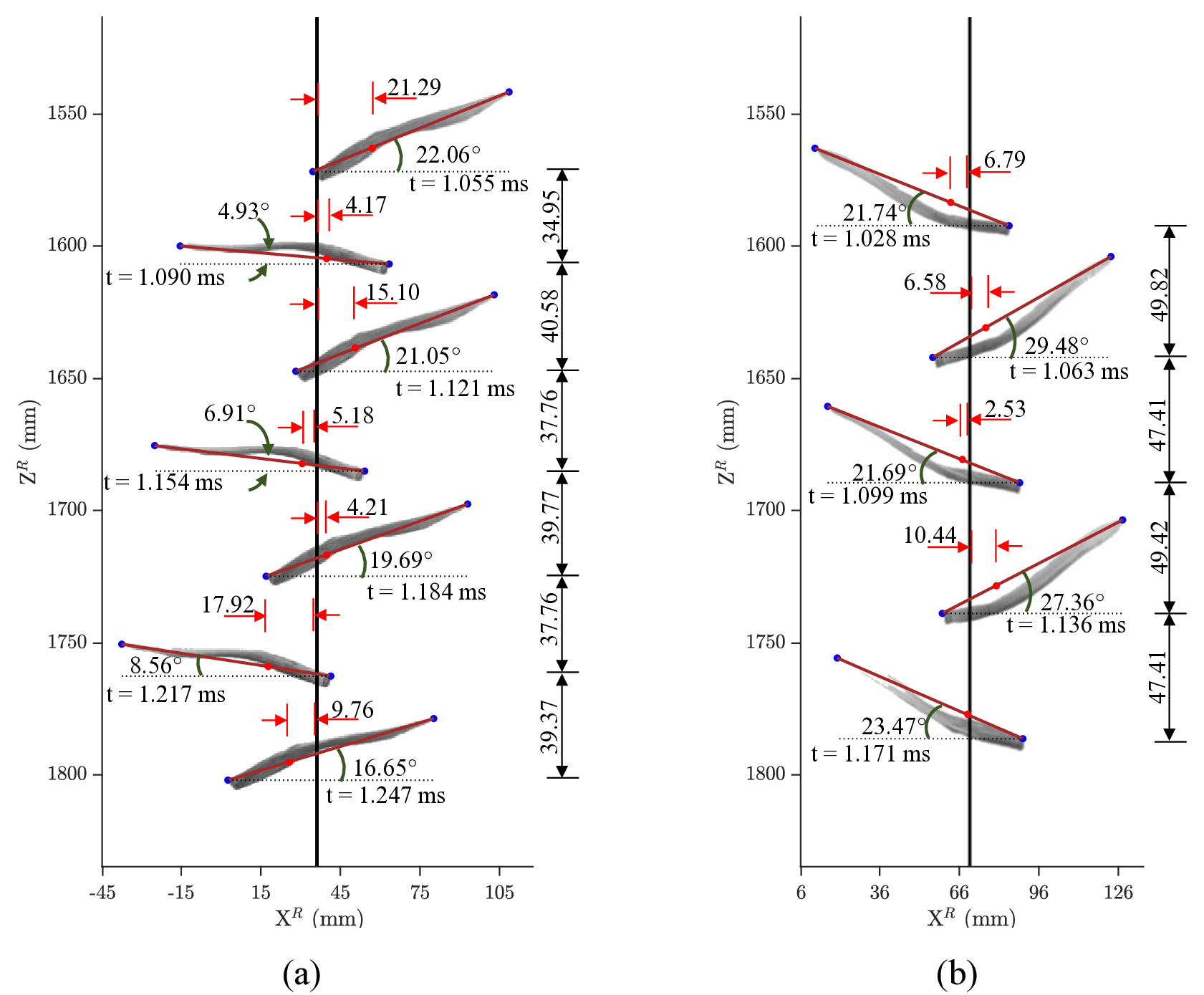}
		\caption{Measured steady-state parameters for the Buddha coconut samara: (a) BD1- clockwise rotation (run 11) and (b) BD2- anticlockwise rotation (run 01). The wingtip pitch (per half rotation) and the radius of precession are reported in millimetres (mm).}
		\label{cone_buddha_overlays}
	\end{figure*}

	For the first set of analyses, representative image overlays of mahogany (MD2) and Buddha coconut (BD1) samaras are shown in Fig.\ref{Mahogany_Buddha_overlays}.
	The MD2 samara (MD2\_Run19\_Anti-clockwise), released in the root-tip orientation, undergoes counterclockwise rotation (Fig.\ref{Mahogany_Buddha_overlays}-a), while the BD1 samara (BD1\_Run11\_Clockwise), released in the wing-tip orientation, exhibits clockwise rotation (Fig.\ref{Mahogany_Buddha_overlays}-b). The consistent repetition of two to three rotations within the field of view confirms that both samaras attained steady-state motion. To perform a comparative analysis for the same sample under opposite directions of rotation, the BD1 samara (BD1\_Run01\_Anti-clockwise) with the root-tip release orientation and the MD2 samara (MD2\_Run24\_Clockwise) with the wing-tip release orientation were considered for further analysis. 
	
	To extract the left and right extreme orientations, the projected area-time plot was used for the mahogany samara during anti-clockwise rotation, whereas for all other three cases, the projected span-time plot was used based on the observed periodicity of the respective signals. The variations of projected area and projected span with time for the four different cases are shown in Fig.\ref{projected_lengths_width}. The local minima in the projected area plots and the local maxima in the projected span plots correspond to the left and right extreme orientations of the samara, respectively. These extrema were extracted and used to compute the steady-state parameters.  The image overlays corresponding to the left and right extreme orientations for the mahogany samaras, under both clockwise and anti-clockwise rotations, are shown in Fig.\ref{Cone_mahogany_overlays} (a--b). Similarly, the overlays for the Buddha coconut samples are presented in Fig.\ref{cone_buddha_overlays} (a--b). The steady-state parameters measured from these extreme orientations, such as the radius of precession, coning angle, wingtip pitch, and the time instances when the samaras reach the left and right extremes, are annotated on the same plots. This provides a clearer understanding of how these parameters vary as functions of time.
	
	Even without extensive explanation, Fig.\ref{Cone_mahogany_overlays} and Fig.\ref{cone_buddha_overlays} clearly highlight the variation in the different steady-state parameters within the steady state, as well as their differences when the samaras reach steady rotation in either the clockwise or anti-clockwise direction. For the mahogany samaras, the coning angle ranges from \(19.78^\circ\) to \(27.44^\circ\) during anti-clockwise rotation, whereas for clockwise rotation it lies within a much narrower band of \(12.81^\circ\) to \(13.80^\circ\). Similarly, for the Buddha coconut samaras, the coning angle during anti-clockwise rotation lies in the range of \(4.93^\circ\) to \(22.06^\circ\), while for clockwise rotation it lies in the range of \(21.69^\circ\) to \(29.48^\circ\). The consistent variation across the dataset confirms that the coning angle is not constant during steady-state descent. Furthermore, the steady-state values of the coning angle differ between the two rotational directions, with anti-clockwise rotation showing a higher range than clockwise rotation for the mahogany samaras, whereas the Buddha coconut samaras display the inverse behaviour.
	
	\begin{figure*}[!h]
		\centering
		\includegraphics[width = 0.8\textwidth]{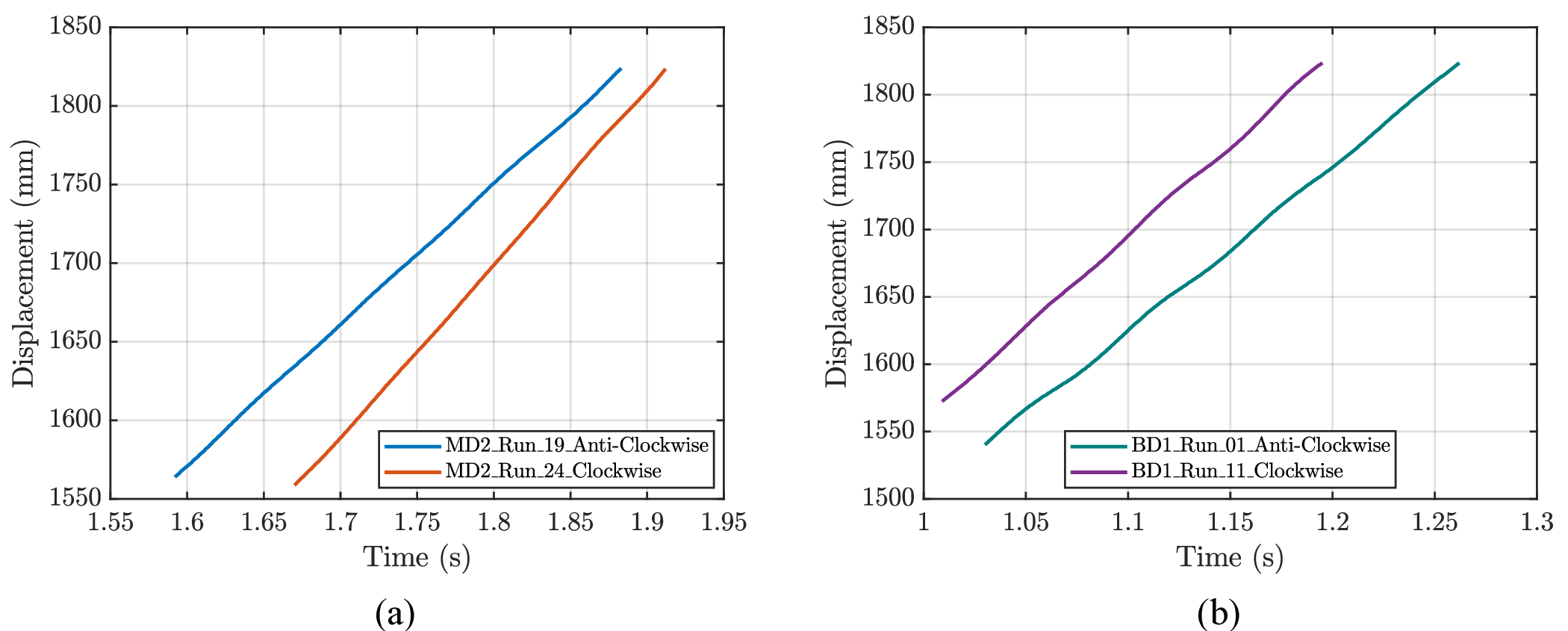}
		\caption{Displacement vs.\ time for (a) Mahogany samara MD2—anticlockwise rotation (run 19) and clockwise rotation (run 24), and (b) Buddha coconut samara BD1 - clockwise rotation (run 11) and anticlockwise rotation (run 01).}
		\label{dispalcement}
	\end{figure*}
	
	\begin{figure*}[!h]
		\centering
		\includegraphics[width = 0.8\textwidth]{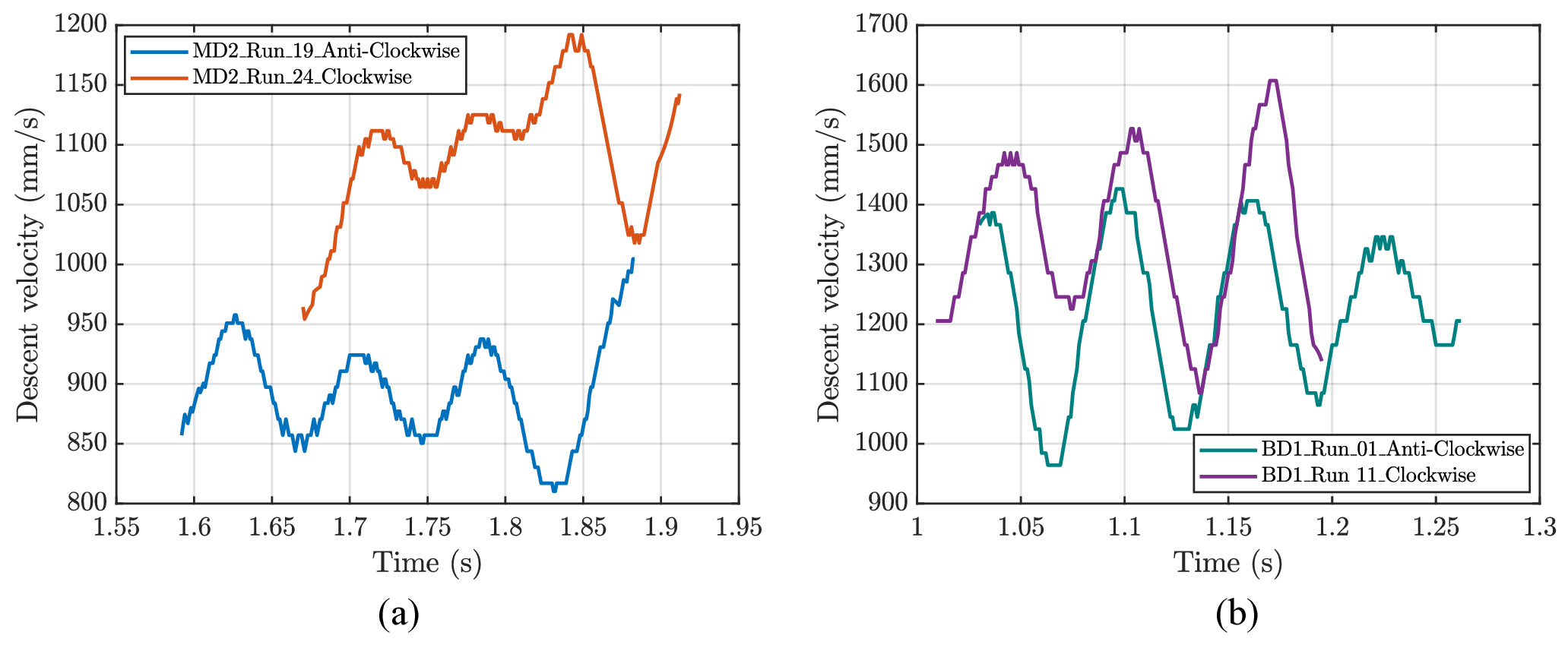}
		\caption{Instantaneous velocity vs.\ time for (a) mahogany samara MD2 -anticlockwise rotation (run 19) and clockwise rotation (run 24), and (b) Buddha coconut samara BD1 - clockwise rotation (run 11) and anticlockwise rotation (run 01).}
		\label{velocity}
	\end{figure*}

	An additional noteworthy observation is the rate of change of the coning angle, which plays an important role in simplifying the angular momentum equations governing spinning samaras. Although continuous measurements would be required for an exact determination, an approximate range can be inferred from the present experimental data. The rate of change of the coning angle is estimated by evaluating the difference in cone angle values at consecutive time instances and dividing by the corresponding time interval. For the Buddha coconut samaras, the rate of change of the coning angle in both clockwise and anti-clockwise rotational directions is found to lie within the range of approximately $100^\circ/\mathrm{s}$ to $500^\circ/\mathrm{s}$. In contrast, the mahogany samaras exhibit a lower range, from about $5^\circ/\mathrm{s}$ to $146^\circ/\mathrm{s}$, based on the temporal variations shown in Fig.\ref{Cone_mahogany_overlays} and Fig.\ref{cone_buddha_overlays} of the same plot. These findings are consistent with the trends reported by Ulrich and Pines~\cite{Ulrich2008PlanformSamara}, reinforcing the conclusion that the coning angle is neither small nor constant, and that its time rate of change must be accounted for when formulating the governing equations for samara dynamics.
	
	The angular velocity of the samaras, calculated from the time required to complete one full rotation, appears to be nearly constant and is observed consistently for both samples. The observed variation in the time taken to complete one rotation is less than 1~ms, which corresponds to the frame interval of the camera. An interesting observation, however, is that the time taken to complete half a rotation is not uniform. The duration from the left extreme orientation to the right extreme differs from that of the return motion. Despite this asymmetry within a single cycle, the total time for one complete rotation remains almost constant. This makes it difficult to assert whether the instantaneous angular velocity is strictly constant. However, the consistency of the full-rotation period indicates that the steady-state approximation for the rotational velocity seems to be valid for the overall motion. As expected, the angular velocity values differ between clockwise and anti-clockwise rotations. For the mahogany samaras (MD2), the angular velocity in the anti-clockwise direction is higher than in the clockwise direction, whereas the opposite trend is observed for the Buddha coconut (BD1) samaras.

	The location of the CM (red dots) describes the precession motion of the samaras. If the CM lies along a straight vertical line, no precessional motion occurs. However, if it deviates from this line, the samara undergoes precession. For all four cases examined, a clear precession motion is observed with respect to the approximated vertical precessional axis (black line), as indicated by the varying lateral distance from this axis. It must be noted that accurately quantifying the radius of precession is challenging because precise tracking of the precession axis is not possible with the present experimental setup. The values computed using an approximate method are annotated in the corresponding figures. The primary conclusion that can be drawn from these plots is that a definite precession motion exists in the samaras within the span range considered in this study. To date, the radius of precession and the precession rate have not been quantitatively determined for real samaras in any previous work.
	
	The pitch of the samaras, defined as the vertical distance swept during one rotation, is estimated by tracking the root-tip point used in the cone-angle measurements. Similar to the rotational velocity, the half-pitch values vary between the two halves of a rotation, whereas the pitch for one complete rotation remains nearly constant, with variations of less than 2~mm for all cases except the clockwise rotation of the mahogany sample, which shows a variation of about 6~mm. It should also be noted that variations in the coning angle can influence the measured pitch, as the root tip undergoes relative motion with respect to the CM.  The pitch of the samaras is a function of both their rotational velocity and descent velocity. Since the pitch per rotation remains nearly constant, while the half-pitch varies, this suggests that the descent velocity may vary periodically within a rotation to maintain a consistent rotation period and net pitch. This hypothesis needs to be examined further using full 3D motion tracking in future studies.

	For all four cases, the trace of the bottommost point of the samaras follows an approximately linear trend, with slight undulations observed for the Buddha coconut samples. The displacement-time plots and their corresponding instantaneous velocity profiles for these cases are shown in Fig.\ref{dispalcement} and \ref{velocity}. It should be emphasized that the coning motion of the spinning samaras significantly influences the trajectory of the bottommost point, much like its effect on the estimation of the wingtip pitch. As a result, the instantaneous velocity measurements exhibit noticeable noise, and a moving-average filter is applied to the velocity data to reduce this effect. For both samples, the clockwise rotation appears to descend at a higher velocity than the anti-clockwise rotation. The nearly periodic pattern observed in the velocity–time curve (Fig.\ref{velocity}) of the bottommost point resembles the trends reported by Varshney et al. \cite{Varshney2011TheMotion}. However, this behavior is primarily influenced by the local variations in the cone angle. Therefore, concluding that the descent velocity itself varies significantly with time would be an overinterpretation of the present data.

	\subsection{Comprehensive parameter variations across the dataset}
	
	\begin{figure*}[!h]
		\centering
		\includegraphics[width = 0.75\textwidth]{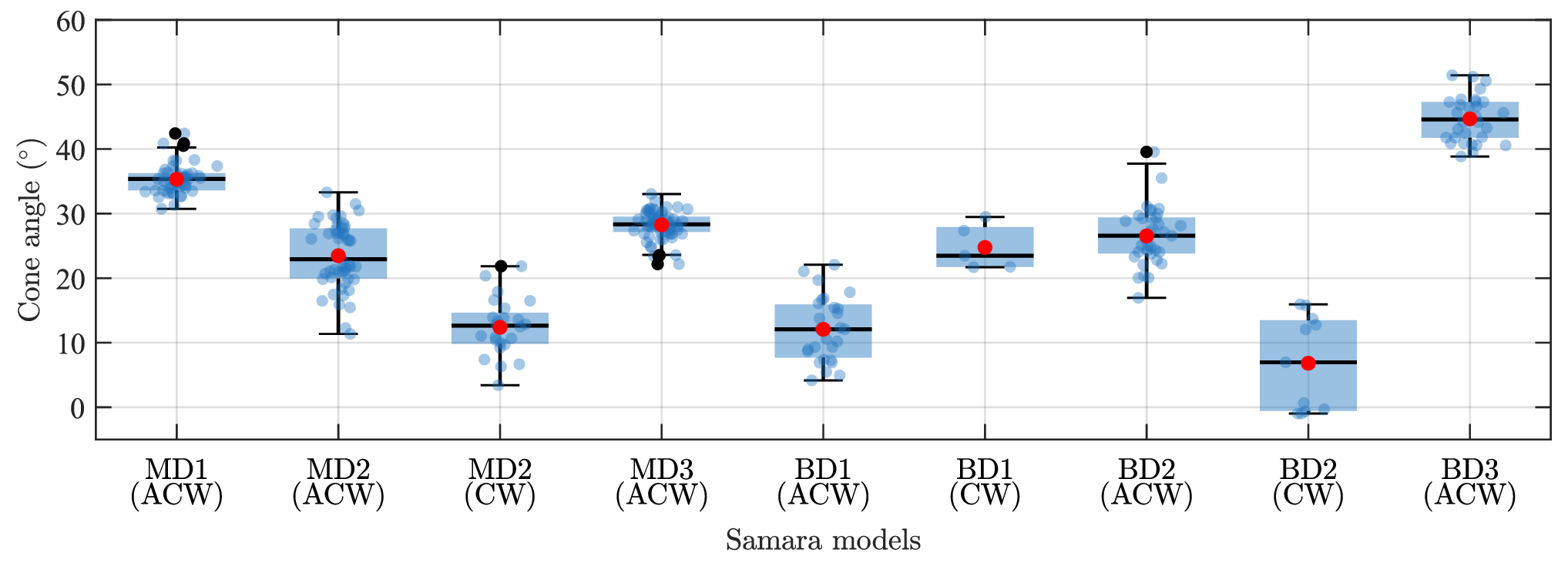}
		\caption{Box plot of cone angle for mahogany samaras (MD1, MD2, and MD3) and Buddha coconut samaras (BD1, BD2, and BD3) across all runs. The text below each sample ID indicates the rotation direction: clockwise (CW) or anticlockwise (ACW).}
		\label{cone_angle_scater}
	\end{figure*}
	
	\begin{figure*}[!h]
		\centering
		\includegraphics[width = 0.75\textwidth]{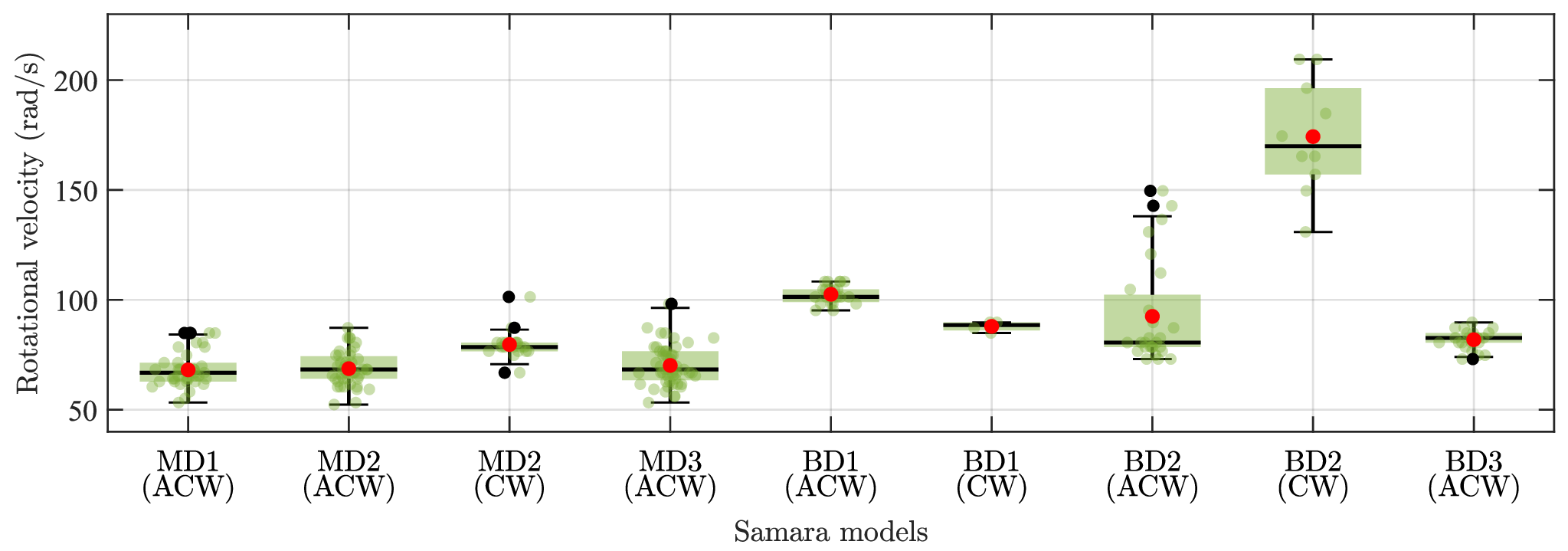}
		\caption{Box plot of rotational velocity (calculated per half rotation) for mahogany samaras (MD1, MD2, and MD3) and Buddha coconut samaras (BD1, BD2, and BD3) across all runs. The text below each sample ID indicates the rotation direction: clockwise (CW) or anticlockwise (ACW).}
		\label{roataional_velocity_scatter}
	\end{figure*}

	\begin{figure*}[!h]
		\centering
		\includegraphics[width = 0.75\textwidth]{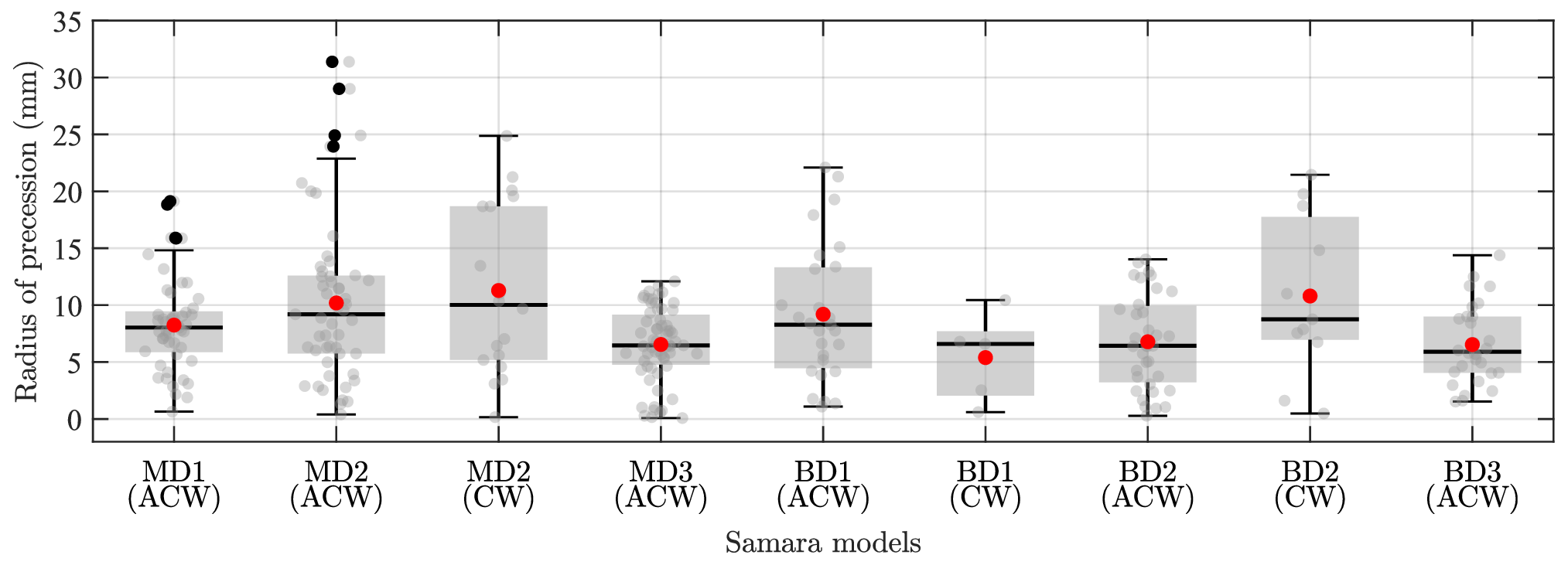}
		\caption{Box plot of radius of precession for mahogany samaras (MD1, MD2, and MD3) and Buddha coconut samaras (BD1, BD2, and BD3) across all runs. The text below each sample ID indicates the rotation direction: clockwise (CW) or anticlockwise (ACW).}
		\label{precession_scatter}
	\end{figure*}

	\begin{figure*}[!h]
		\centering
		\includegraphics[width = 0.75\textwidth]{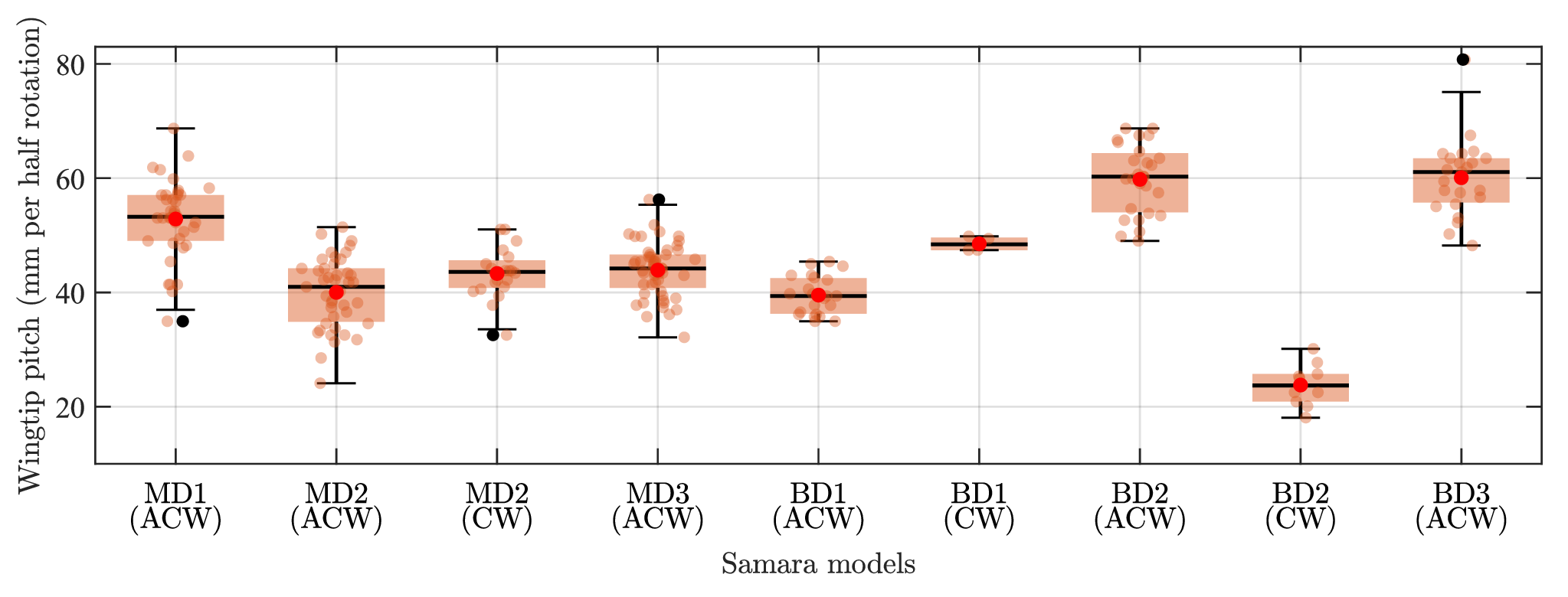}
		\caption{Box plot of wingtip pitch (per half rotation) for mahogany samaras (MD1, MD2, and MD3) and Buddha coconut samaras (BD1, BD2, and BD3) across all runs. The text below each sample ID indicates the rotation direction: clockwise (CW) or anticlockwise (ACW).}
		\label{wingtip_pitch_sc}
	\end{figure*}

	\begin{figure*}[!h]
		\centering
		\includegraphics[width = 0.75\textwidth]{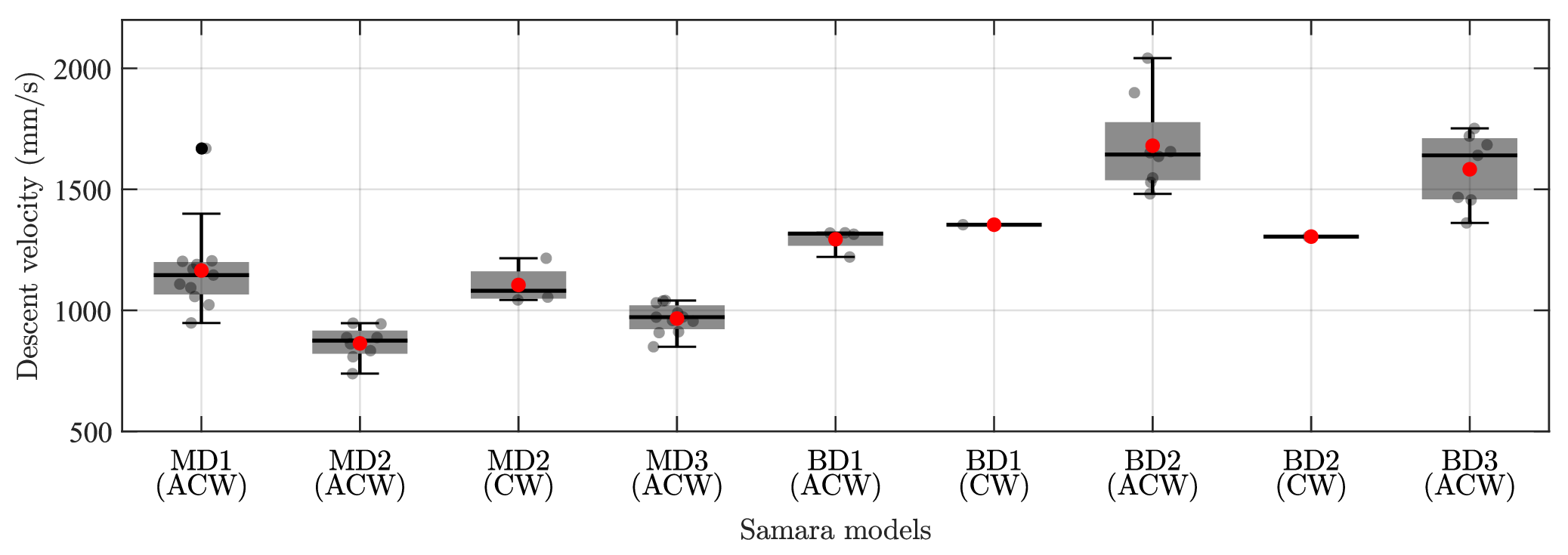}
		\caption{Box plot of descent velocity for mahogany samaras (MD1, MD2, and MD3) and Buddha coconut samaras (BD1, BD2, and BD3) across all runs. The text below each sample ID indicates the rotation direction: clockwise (CW) or anticlockwise (ACW).}
		\label{deescent_velcoity_scatter}
	\end{figure*}

	Similar to the four representative cases discussed earlier, the results were extracted for the remaining drop experiments and plotted to statistically determine whether the values of each parameter remain constant or vary for any of the steady-state quantities across all samples. Only those runs in which the samaras fully entered the steady state were included in this analysis. In the case of the Buddha coconut samaras, the transition length is considerably larger than that of the mahogany samaras, resulting in few drops exhibiting pre--steady-state behavior within the field of view for different release orientations. The periodicity of the projected-span or projected-area plots was examined to confirm whether the samara had fully reached steady state. Only data satisfying this condition were used for the statistical evaluation.
	
	Previous studies \cite{Lee2016NumericalSeed,Lee2016EffectPalmatum} have suggested a correlation between specific drop orientations and the resulting direction of rotation. In contrast, the present observations show that even identical release orientations can lead to both clockwise and anti-clockwise rotations, likely due to the complex dynamics during the transition phase. Consequently, the rotational direction for each sample is determined individually and incorporated into the statistical analysis as described below. In this regard, the MD2 samples of the mahogany samaras exhibited both clockwise and anti-clockwise rotation, while among the Buddha coconut samaras, this behaviour was observed in BD1 and BD2. For clarity in the statistical plots, the direction of rotation is indicated below each model label, with anti-clockwise rotation denoted as (ACW) and clockwise rotation as (CW).
	
	The statistical analysis of the extracted steady-state kinematic parameters was performed using box plots, a standard tool for visualizing the distribution of data within a sample. Box plots are employed because they clearly represent data variability, highlight robust measures of central tendency, allow straightforward identification of outliers, and facilitate effective comparison across multiple runs. In a box plot, the central box represents the interquartile range (IQR), defined by the 25th and 75th percentiles. The red dot inside the box denotes the median, while the horizontal black line indicates the mean value of the data set. The whiskers extend to the most extreme data points that lie within 1.5 $\times$ IQR from the quartiles. The scattered light-colored points represent the individual measurements for each run, with their shade varying across cases to match the corresponding data set. Data points lying beyond the whisker limits are classified as outliers and are shown as distinct black markers. This representation enables clear visualization of the data spread, central tendency, consistency across trials, and the presence of outliers for each samara model. The box plots for rotational velocity, half-wingtip pitch, radius of precession, coning angle, and descent velocity are shown in Fig.\ref{cone_angle_scater}-\ref{deescent_velcoity_scatter}. The descent velocity was obtained from the slope of a linear fit to the displacement-time data, and this averaged value was used for the statistical analysis rather than instantaneous measurements. The wingtip pitch and rotational velocity were computed for each half-rotation in order to quantify their variation across runs.

	Examination of the resulting plots shows noticeable differences in box widths, whisker lengths, and the distribution of scattered points across all cases. These variations clearly indicate that none of the measured kinematic parameters remain constant during the steady state of the samaras. The cone angle (Fig.\ref{cone_angle_scater}) exhibits a distinct spread in its values, reflecting fluctuations in the coning motion across runs. Likewise, the descent velocity (Fig.\ref{deescent_velcoity_scatter}) shows considerable dispersion, confirming that the falling speed varies between drops rather than settling into a fixed value. The rotational velocity, calculated per half rotation, shows significant variability, indicating noticeable changes in the spin rate between successive half rotations (Fig.\ref{roataional_velocity_scatter}). The radius of precession (Fig.\ref{precession_scatter}) spans a wide range, demonstrating substantial differences in the precessional motion among samples. Finally, the wingtip pitch (per half rotation) shows marked variation within each case (Fig.\ref{wingtip_pitch_sc}) . Together, these observations underscore the dynamic nature of the flight behaviour of the samaras, with each parameter exhibiting measurable variation.

	
	\subsection{Constraints of the steady-state approximation and their implications}
	
	Thus, from the above results and discussion, we conclude that in the steady state, the descent velocity, coning angle, and motion of the CM of the samaras are not as simple as previously assumed in the work of Azuma and Yasuda \cite{Azuma1989FlightSeeds}. This reveals an important point that although many studies (as listed in table~\ref{tab:samara_analysis_summary}) have simplified the governing equations based on these assumptions, such simplifications may miss the actual complexity seen in real samara flight. Even though the assumptions proposed by Azuma and Yasuda \cite{Azuma1989FlightSeeds} greatly simplify the governing equations, they may fail to fully capture the actual physics governing the motion. This highlights the need to solve the complete set of governing differential equations without relying on such simplifications for future studies

	\begin{figure}[!h]
		\centering
		\includegraphics[width=0.5\textwidth]{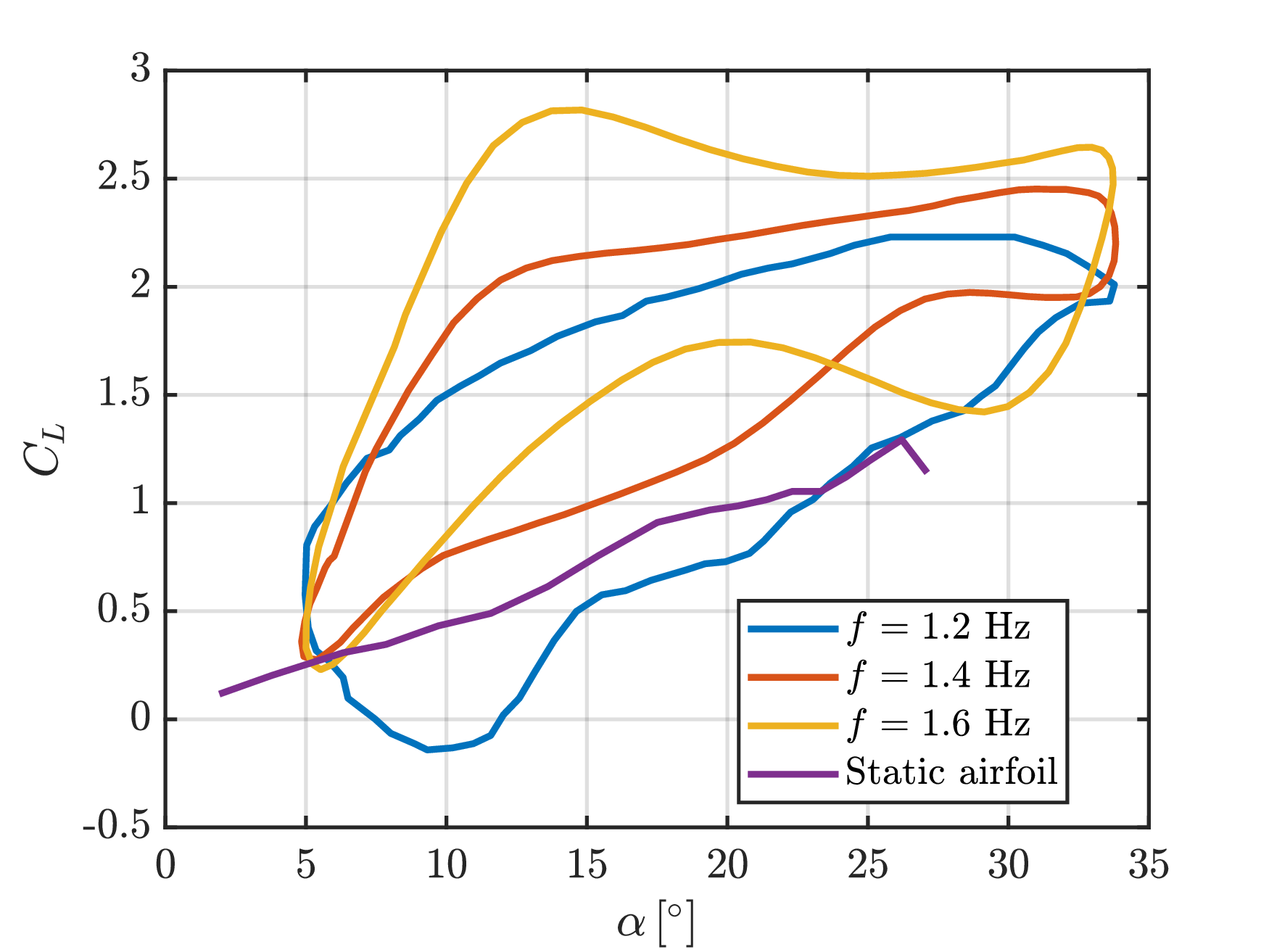}
		\caption{Comparison of lift coefficient (\(C_L\)) versus angle of attack (\(\alpha\)) for static and dynamically pitching airfoils at Reynolds number \(Re = 1000\). The dynamic case involves a pitch-down motion with a mean angle of \(\alpha_0 = 20^\circ\) and amplitude \(\alpha_{\text{amp}} = 15^\circ\). Data reconstructed from Liu et al. \cite{Liu2012NumericalPerturbation}}. 
		\label{fig:dynamic_static_pitchdown}
	\end{figure}

	On the other hand, the combined effects of pitching, coning, and precession have a substantial influence on the aerodynamics of spinning samaras. For instance, the study by Liu et al. \cite{Liu2012NumericalPerturbation} demonstrated that a continuously pitching airfoil at $Re = 1000$ exhibited up to 80\% higher lift compared to a stationary airfoil, as illustrated in Fig.\ref{fig:dynamic_static_pitchdown}. This is a significant finding and suggests that the dynamic motions of samaras could strongly affect lift generation and the behavior of the LEV \cite{Lentink2011Leading-EdgeOf,Yogeshwaran2021AerodynamicsModel,G2023InvestigationPIV,Salcedo2013StereoscopicPIV}, which warrants further investigation.
	
	The current methodology offers only qualitative insights into the steady-state parameters of spinning samaras, as it relies on 2D captures of their inherently 3D motion using high-speed cameras. Ideally, the full 3D trajectory of the samaras should be tracked using a dual high-speed camera setup to obtain detailed quantitative measurements. Although considerable efforts were made in this study to reconstruct the 3D motion using two synchronized cameras, the limited spatial resolution and the small size of the samaras presented significant challenges, making accurate 3D trajectory reconstruction difficult to accomplish.
	
	The DNS studies conducted by Lee et al. \cite{Lee2016NumericalSeed}, which captured the variation in pitching and coning angles, were not validated using experimental data. In fact, there is currently no experimental dataset available for natural samaras that fully tracks their complete 3D motion, except for the partial efforts by Varshney et al. \cite{Varshney2011TheMotion}. To properly validate DNS simulation results, a comprehensive experimental investigation of the full 3D motion is essential, involving accurate measurement of all relevant parameters and comparison with simulation outputs. To visualize the absolute motion of samaras, the trajectories obtained in both the previous and present studies were simulated for conceptual understanding. These visualizations clearly highlight the complex nature of samara flight paths, emphasizing the importance of conducting detailed experimental studies in the future.

	\section{Hypothesized Steady-State Dynamics of Single-Winged Samaras}

	\begin{table*}[!h]
		\centering
		\caption{Kinematic parameters used to trace the trajectory of samara motion.}
		\label{literature_table_steady_state}
		\begin{tabular}{@{\hskip 5pt} c @{\hskip 24pt} l @{\hskip 12pt} c @{\hskip 12pt} l @{\hskip 5pt}}
			\toprule
			\textbf{S.No.} & \textbf{Parameter} & \textbf{Value} & \textbf{Reference} \\
			\midrule
			1 & Descent velocity & 0.9 m/s & present study \\
			2 & Rotation velocity about own axis & 62.83 rad/s & present study  \\
			3 & Precession velocity  & 30 rad/s & \cite{Varshney2011TheMotion} \\
			4 & Coning angle  &  35$^\circ$ to 55$^\circ$&  present study\\
			5 & Pitching angle & -5$^\circ$ to 5$^\circ$  & \cite{Lee2016NumericalSeed} \\
			6 & Radius of precession& 20 mm  & present study \\
			7 & Coning rate & 164 rad/s &  \cite{Lee2016NumericalSeed}\\
			8 & Pitching rate & 164 rad/s & \cite{Lee2016NumericalSeed} \\
			\bottomrule
		\end{tabular}
		\label{tab:samara_parameters}
	\end{table*}
	
	\begin{figure*}[!h]
		\centering
		\includegraphics[width = 0.75\textwidth]{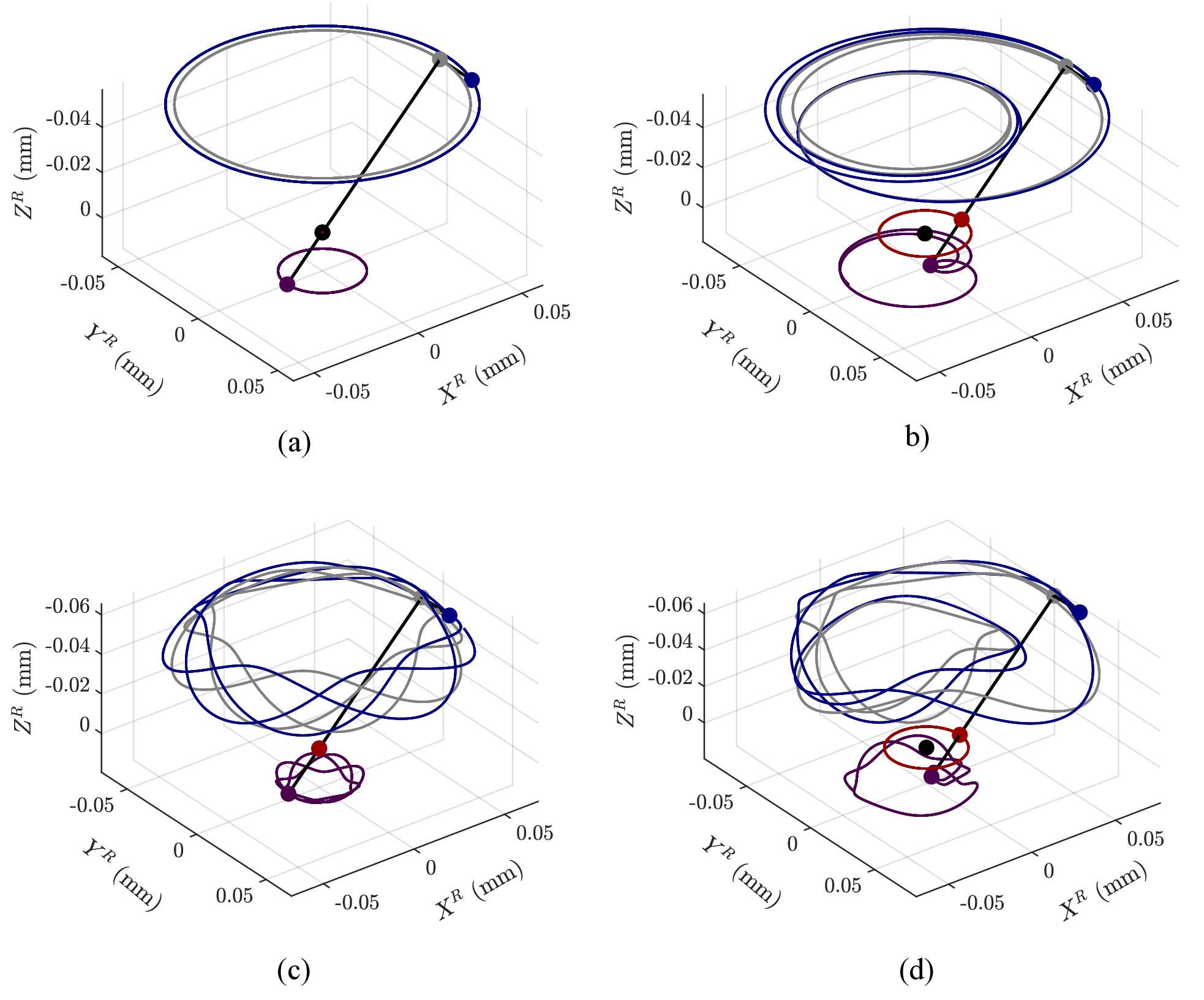}
		\caption{Trajectory of the samaras under the assumption of zero descent velocity. The red curve shows the trace of the CM, the gray curve shows the trace of the wingtip mid-chord point, the blue curve shows the trace of the wingtip trailing edge, and the violet curve shows the trace of the root tip. (a) depicts the motion of the samara under a constant coning angle without precession or pitching. (b) adds precession to the motion in (a). (c) adds coning motion to (a), and (d) combines the motions in (a), (b), and (c) along with pitching dynamics.}
		\label{Ideal_trajec_zero_descent}
	\end{figure*}
	
	\begin{figure*}[!h]
		\centering
		\includegraphics[width = 0.9\textwidth]{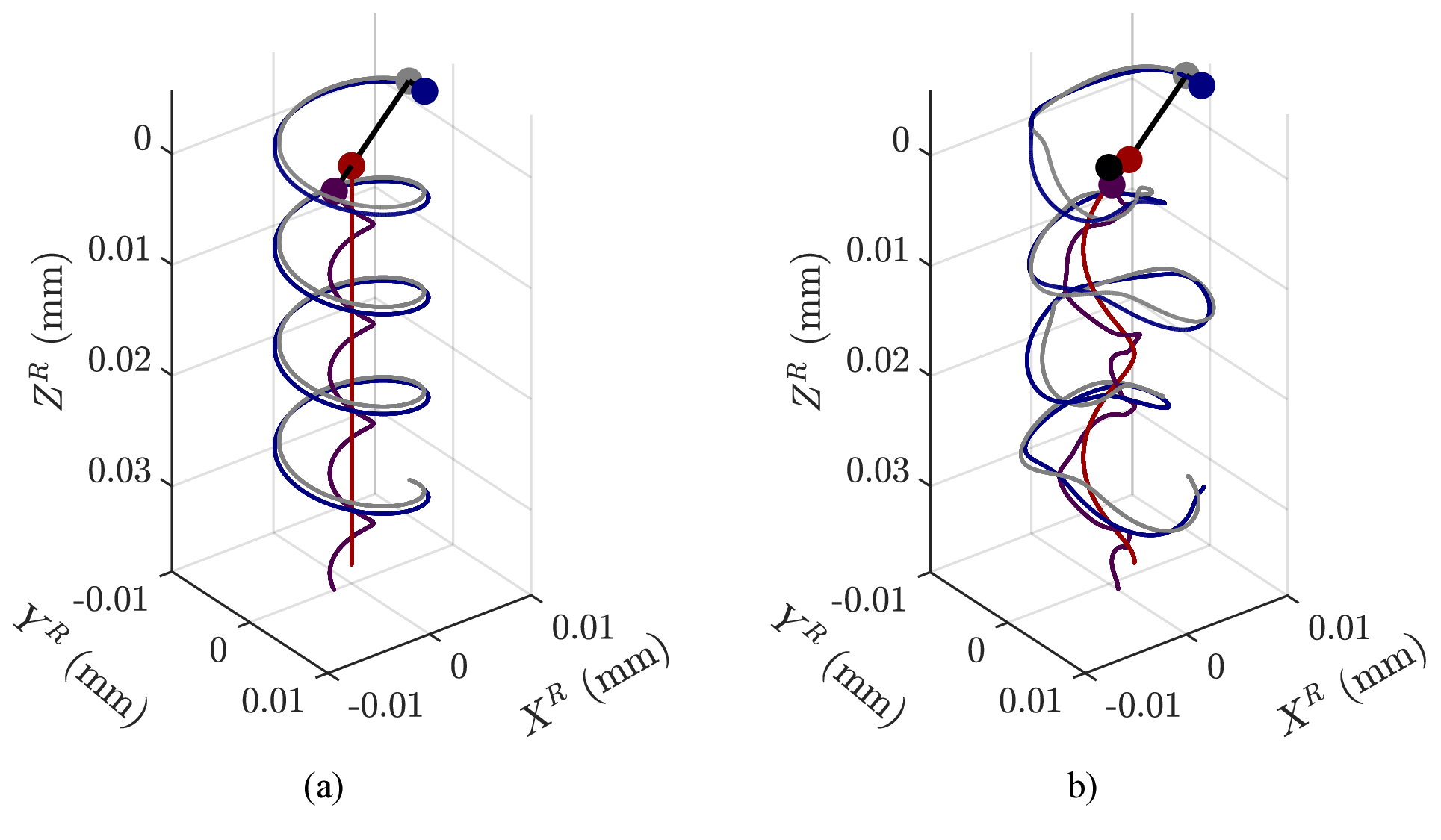}
		\caption{Trajectory of the samaras with a descent velocity of 0.9 m/s for cases (a) and (d) in figure~\ref{Ideal_trajec_zero_descent}.}
		\label{Ideal_trajec_descent}
	\end{figure*}
	
	With the results obtained from the present study, the trajectory of the samara in its steady-state motion can be traced to gain a better understanding of its kinematic behavior. Four characteristic points were identified on the natural samaras to represent their steady-state trajectory: the CM, the leading edge of the wingtip, the trailing edge of the wingtip, and the root tip, which lies along the line joining the leading edge of the wingtip and the CM. The geometric parameters derived from the present study yielded a representative span of 80 mm, a CM located 30\% from the root tip, and a wingtip chord of 20 mm, which were subsequently used to trace the trajectory of the samara.

	The trajectory was constructed using a few approximations of the observed motion, since the variation of each parameter is not fully available from the literature or the present study. The CM follows a helical path about the vertical ($Z^R$) axis, corresponding to the precession of the samara. The descent velocity is assumed to remain constant, and the rotational velocity of the samara about its CM is also treated as constant, although both exhibit variations in the actual case. The leading edge of the wingtip undergoes a periodic coning motion with respect to the CM, while the trailing edge experiences a periodic variation in pitch angle relative to the leading edge. The cone angle was constrained to the range of $35^{\circ}$ to $55^{\circ}$, which is the upper limit observed in the present study and was selected for visualization purposes without any loss of generality. Although this representation appears to describe an exact motion, results from Lee \cite{Lee2016NumericalSeed} indicate that both the coning and pitching angles can exhibit non-periodic variations. The parameters used to trace the steady-state trajectory of the samara, together with their corresponding sources, are presented in the table~\ref{tab:samara_parameters}.
	
	To capture and visualize the complete complexity of the motion of the samara, the trajectory is plotted in two phases. In the first phase, the trajectory is constructed in the reference frame of the samara, assuming zero descent velocity. Four configurations were simulated, starting from a simplified trajectory proposed by Azuma and Yasuda \cite{Azuma1989FlightSeeds} (configuration 1), followed by the inclusion of precession (configuration 2), variation in the coning angle (configuration 3), and finally a combined motion incorporating pitching, coning, and precession (configuration 4). These trajectories are illustrated in Fig.\ref{Ideal_trajec_zero_descent}. In the second phase, the descent velocity component was introduced to configurations 1 and 4 to obtain the corresponding actual trajectories, as shown in Fig.\ref{Ideal_trajec_descent}.
	
	This two-phase visualization provides a clearer representation of the inherent complexity in the steady-state motion of spinning samaras and aids in effectively visualizing their kinematics. These results distinctly illustrate how the trajectory of what may appear to be a simple falling object can, in reality, exhibit intricate and highly coupled motion, as observed in nature. Furthermore, whether employing DNS or BEMT, such trajectories represent the type of motion that future studies can be expected to model and simulate.

	\section{Steady-state assumptions and approximations}
	Although the current results may appear to complicate the simplified frameworks employed in previous studies, they highlight the necessity of incorporating all relevant parameters to obtain a complete and accurate solution. Under such conditions, the formulation becomes mathematically intractable because no valid simplifications or approximations can be applied, necessitating the solution of the complete set of governing equations. However, the observed sinusoidal variations of the cone angle, pitch angle, and three velocity components, together with the linear variation of the rotational angle in the steady-state motion of the samara, lead to a significant simplification of the governing equations. Based on these experimentally validated conditions, the system of ordinary nonlinear differential equations that governs the dynamics of the samara can be reduced to a set of algebraic relations. To provide a preliminary analysis of such solutions, the translational velocity components of the CM are expressed following the experimental results reported by Varshney et al. \cite{Varshney2011TheMotion}, with the simplifying assumption that each velocity component is primarily influenced by a single dominant frequency, although, in reality, multiple frequency components may be present.

	\begin{equation}
		\begin{aligned}
			x^{R}(t) &= r_m \sin(c_1 t + c_2), \\
			y^{R}(t) &= r_m \cos(c_3 t + c_4), \\
			z^{R}(t) &= c_7t + c_5 \sin(c_6 t).
		\end{aligned}
		\label{tranlational}
	\end{equation}
	
	where \( r_m \) denotes the radius of precession of the CM and \( c_i \) are constant parameters. Similarly, the angular motions of the samara are expressed as harmonic functions of time \cite{Lee2016NumericalSeed,Ulrich2008PlanformSamara},
	\begin{equation}
		\begin{aligned}
			\theta_{LR}(t) = A_1 \cos(b_1 t + C_1), \\
			\phi_{LR}(t) = A_2 \cos(b_2 t + C_2), \\
			\psi_{LR}(t) = m t + \psi_0.
			\label{roatiaonl}
		\end{aligned}
	\end{equation}
	with amplitudes \( A_1, A_2 \), frequencies \( b_1, b_2 \), phases \( C_1,C_2 \)  and a steady spin rate \( m \).

	By incorporating equations (\ref{tranlational}) and (\ref{roatiaonl}) into the governing differential equations(~\ref{eq_newtons law},~\ref{eq_eulers law},~\ref{eq:navier_stokes},~\ref{eq:continuity}), the system of nonlinear ordinary differential equations reduces to a set of algebraic equations, which can be solved more easily than those obtained using the previous assumptions and approximations in earlier studies. Note that these simplifications apply only to the steady-state motion of the samara, whereas during the transition phase, the kinematics differ considerably. The results reported by Ulrich and Pines \cite{Ulrich2008PlanformSamara}, Lee \cite{Lee2016NumericalSeed}, and Varshney et al. \cite{Varshney2011TheMotion} suggest an inherent relationship among the velocity components, manifested through their frequency and phase variations. In particular, the works of Ulrich and Pines \cite{Ulrich2008PlanformSamara} show that the pitch and cone angles oscillate at similar frequencies but exhibit a consistent phase difference relative to each other.
	
	If such relationships can be established more precisely, the number of unknowns in the governing equations can be significantly reduced, thereby simplifying the overall formulation. Consequently, the motion could be solved more efficiently using either BEMT or DNS by considering only the steady-state phase, provided that the geometric parameters and the external forces and moments are well defined. Establishing these relationships and defining these parameters will constitute an important part of future work.

	\section{Conclusions}
	The present study reveals that key steady-state kinematic parameters of spinning samaras, such as descent velocity, coning angle, pitching motion, and precessional movement, exhibit significant temporal variations, contradicting the steady and simplified behaviour assumed in earlier models. These findings highlight that such assumptions, while mathematically convenient, fail to capture the inherent complexity of real samara flight and may lead to incomplete physical interpretations. At the same time, the experimentally observed sinusoidal variations in cone angle, pitch angle, and translational motion, together with the nearly linear rotation rate, offer an alternative path for meaningful simplification. Representing these motions through harmonic functions enables the governing nonlinear differential equations to be reduced to a more tractable algebraic form without overlooking essential physics. Furthermore, the extracted trajectories of the samaras reveal the inherent difficulty of visualizing and predicting their complex motion using either computational or experimental approaches alone. In the absence of complete three-dimensional experimental datasets for natural samaras, detailed kinematic measurements are essential for the validation of high-fidelity numerical models such as DNS and BEMT. In summary, this study demonstrates that a realistic description of samara aerodynamics must explicitly account for the natural variability of kinematic parameters. At the same time, their underlying oscillatory structure can be exploited to develop modeling frameworks that are both physically accurate and computationally efficient.
	
	\section{Appendix A}
	\subsection{Cone angle calculations}
	\begin{figure}[!h]
		\centering
		\includegraphics[width=0.3\textwidth]{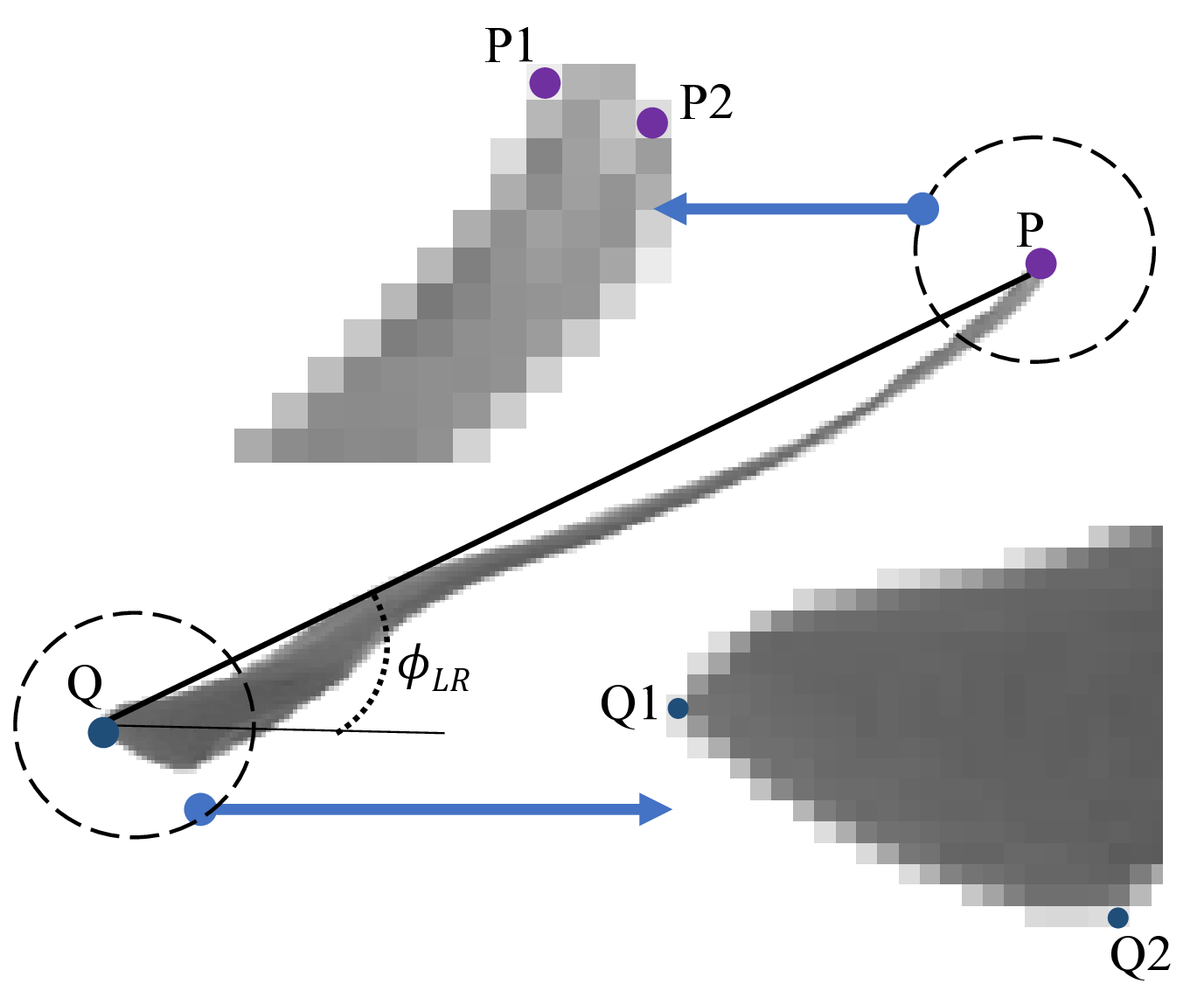}
		\caption{Identification of the wingtip and root–tip points used for coning angle calculation}
		\label{cone_enlarged}
	\end{figure}
	
	Fig.\ref{cone_enlarged} illustrates an enlarged view of the mahogany samara corresponding to the right extreme orientation. Both the top and bottom edges are magnified to highlight various possible points that can be used for calculating performance parameters, particularly for the coning angle at the left and right extreme orientations. As shown, the bottom edge has two possible points (Q1 and Q2), while the top edge has two options (P1 and P2). The existing literature does not provide a clear guideline on which specific points should be selected for computing the coning angle. The choice of points can lead to variations of up to \(5.46^\circ\), as shown in Table~\ref{cone_angle_different_values}. However, this level of variation does not significantly affect the overall results. Therefore, for consistency, the points P2 and Q1 are used for calculating the coning angle across all identified extreme positions.

	\begin{table}[!htbp]
		\centering
		\caption{Angle calculations from pixel coordinates}
		\label{cone_angle_different_values}
		
		\setlength{\tabcolsep}{6pt}
		\renewcommand{\arraystretch}{1.25}
		\footnotesize
		
		\begin{tabular}{|
				p{0.6cm}|
				p{0.8cm}|
				p{0.8cm}|
				p{2.2cm}|
				p{1.2cm}|}
			\hline
			\textbf{Point} &
			\textbf{x}\newline (pixel) &
			\textbf{y}\newline (pixel) &
			\textbf{Points considered} &
			\textbf{Angle}\newline ($^\circ$) \\
			\hline
			
			P1 & 459 & 114 & P1, Q1 & 26.47 \\
			\hline
			P2 & 462 & 115 & P2, Q1 & 25.91 \\
			\hline
			Q1 & 256 & 215 & P1, Q2 & 31.37 \\
			\hline
			Q2 & 277 & 225 & P2, Q2 & 30.74 \\
			\hline
			
		\end{tabular}
	\end{table}

	
	\section*{Acknowledgements}
	{The authors gratefully acknowledge the members of the laboratory for their valuable discussions and assistance, which helped shape this manuscript. The authors sincerely thank Mr S. Abhishek, Technical Assistant in the Department of Aerospace Engineering at the Indian Institute of Science (IISc), for his support in conducting the experiments. The authors also express their appreciation to Mr Adarsh Abraham Basumata, PhD, Department of Computer Science and Automation, Indian Institute of Science (IISc), for his assistance in tracing the trajectory of the samaras.}
	
	
	
	
	
	\section*{Declarations}{Fig.3 and Fig.23 were plotted using data extracted from the works of Varshney et al. (Varshney et al. 2011) and Liu et al. (Liu et al. 2012). The data were digitized using WebPlotDigitizer and subsequently plotted using MATLAB. The inclusion of these two figures strengthens the justification for ruling out several assumptions made in previous studies and highlights the significant effects of these assumptions on aerodynamic behavior.}
	
	

	\end{document}